\documentclass[twocolumn]{jpsj2}

\title{
Continuous-Time Quantum Monte Carlo Method for the Coqblin-Schrieffer Model
}

\author{%
Junya \textsc{Otsuki}\thanks{E-mail address: otsuki@cmpt.phys.tohoku.ac.jp}, 
Hiroaki \textsc{Kusunose}$^1$, Philipp \textsc{Werner}$^2$
and Yoshio \textsc{Kuramoto}
}

\inst{%
Department of Physics, Tohoku University, Sendai 980-8578, Japan\\
$^1$Department of Physics, Faculty of Science, Ehime University, Matsuyama 790-8577, Japan\\
$^2$Columbia University, 538 West, 120th Street, New York, New York 10027, USA
}

\recdate{\today}

\abst{%
An impurity solver based on a continuous-time quantum Monte Carlo method 
is developed for the Coqblin-Schrieffer model. 
The Monte Carlo simulation does not encounter a sign problem for antiferromagnetic interactions, and 
accurately reproduces the Kondo effect.
Our algorithm can deal with an arbitrary number $N$ of local degrees of freedom, becomes more efficient for larger values of $N$, and is hence suitable for models with orbital degeneracy. 
The dynamical susceptibility and the impurity $t$-matrix 
are derived with the aid of the Pad\'e approximation for various values of $N$, and good agreement is found with other methods and available exact results.
We point out that the Korringa-Shiba relation needs correction for a finite value of the exchange interaction.
}

\kword{%
continuous-time quantum Monte Carlo method, Coqblin-Schrieffer model, Kondo model, $t$-matrix, dynamical susceptibility, Pad\'e approximation, Korringa-Shiba relation
}

\begin{document}
\maketitle

\section{Introduction}

Strong correlations among localized and conduction electrons 
lead to the Kondo effect in impurity systems, and 
heavy fermions in a periodic lattice.
Two contrasting  approaches may be used to deal with lattice models theoretically: one is to solve the model on a finite cluster by a method such as exact diagonalization, the other involves the solution of an effective impurity system within the framework of dynamical mean field theory (DMFT)\cite{Georges}. 
The former approach is more suitable for low-dimensional systems, while the latter becomes exact in infinite-dimensional systems. 
For actual heavy fermion materials in three dimensions,  
the DMFT is the simplest approach
which can take local correlations into account.
But even within this approximate framework, perturbative calculations fail 
in the most interesting parameter range, where fierce competition arises between local and inter-site correlations.   
Therefore, numerical approaches are required to reliably solve the effective impurity problem.

In this paper, we present a new impurity solver based on
the continuous-time quantum Monte Carlo method (CT-QMC) for fermion systems, which has been originally proposed by Rubtsov\cite{Rubtsov} et al.
The CT-QMC evaluates the infinite sum of multiple integrals in a perturbation expansion by means of a Monte Carlo procedure. 
The CT-QMC does not require a Trotter decomposition
in contrast with auxiliary field quantum Monte Carlo methods such as Hirsch-Fye. 
While the CT-QMC was first formulated as a perturbation expansion with respect to the Coulomb interaction, 
an alternative expansion with respect to hybridization around the atomic limit
has been developed \cite{Werner}. For the Anderson model, the latter approach does not encounter a sign problem \cite{Yoo}
and empirically, it is found that even more complicated multiorbital and cluster models can be simulated without encountering negative weight configurations. 
Therefore it is a powerful impurity solver for DMFT and its extensions, and has been applied to several models\cite{Werner, Gull, Werner_doping, Haule, Werner_phonon}. 

The CT-QMC solvers \cite{Rubtsov, Werner} are not suitable for very strong correlations. 
In the ``segment" picture of ref.~\citen{Werner}, the simulation of the Anderson model with strong Coulomb repulsion and
a deep local level involves short excursions to high-energy states (short overlapping segments or anti-segments). Without special precautions, the acceptance rate of the Monte Carlo moves will be very low. 
However, the empty and doubly occupied states play an essential role
and give rise to the effective exchange interaction.
Since these high-energy states are easily dealt with by ordinary perturbation theory,
it is reasonable to employ an effective model which eliminates these virtual processes from the start.
It is well known that 
the Coqblin-Schrieffer (CS) model corresponds to such a localized limit of the Anderson model\cite{CS}. 
The purpose of this paper is to formulate the CT-QMC in a form which is applicable to the CS model. 

Our scheme, to be presented below, has the following features:
\begin{itemize}
\item[(i)] As long as interactions are antiferromagnetic, the scheme is free from the minus sign problem. This is consistent with the fact that the CS model with the antiferromagnetic exchange is derived from the Anderson model. 
\item[(ii)] Acceptance probabilities for the random walk are higher compared to the case of the Anderson model. This is because the formulation excludes the charge degree of freedom.
The remaining magnetic exchange processes are sampled efficiently.
\item[(iii)] An arbitrary number $N$ of local states can be dealt with in a simple manner. 
Larger values of $N$ make the computation faster in general, since the number of operators in each component decreases. 
Hence the scheme can easily handle orbital degeneracy. 
\end{itemize}
Since the algorithm does not encounter the minus sign problem, 
highly accurate dynamics can be derived.  The algorithm is therefore especially useful
for large-$N$ systems, where other methods such as exact diagonalization are not practical due to the rapidly increasing number of basis states. 

This paper is organized as follows. 
In the next section, the CT-QMC formalism is presented for the CS model. 
Section 3 discusses the Monte Carlo sampling procedure. 
The algorithm is also applied to the Kondo model
which is discussed in \S4.
Numerical results for static quantities are shown in \S5 for a wide range of $N$. 
Single-particle and two-particle spectra are evaluated with the aid of the Pad\'e approximation in \S6.  
The summary of the paper is given in \S7.

\section{Formulation}

\subsection{Partition function}
In order to present our new algorithm, it is convenient to summarize the general formulation of the CT-QMC\cite{Rubtsov,Werner}. 
The partition function for the Hamiltonian $H=H_0 + H_1$ is given by
\begin{align}
	Z
	= \text{Tr} \left\{ T_{\tau} {\rm e}^{-\beta H_0} \exp \left[- \int_0^{\beta} {\rm d}\tau H_1 (\tau) \right] \right\},
\end{align}
where $H_1(\tau)={\rm e}^{\beta H_0} H_1 {\rm e}^{-\beta H_0}$ is an operator in the interaction picture. 
Expanding the exponent we obtain another expression for the partition function:
\begin{align}
	Z =& \sum_{k=0}^{\infty} \frac{(-1)^k}{k!} \int_0^{\beta} {\rm d}\tau_1 \cdots \int_0^{\beta} {\rm d}\tau_k
	\nonumber \\
	&\times \text{Tr} \{ T_{\tau} {\rm e}^{-\beta H_0} H_1(\tau_1) \cdots H_1(\tau_k) \}.
\label{eq:part_func_gen}
\end{align}
In the CT-QMC, this sum of multiple integrals is sampled stochastically. 

We consider the Coqblin-Schrieffer model with $N$ components\cite{CS}
\begin{align}
	H_0 &= H_{\rm c} + H_f
	= \sum_{\mib{k} \alpha} \epsilon_{\mib{k}} c_{\mib{k} \alpha}^{\dag} c_{\mib{k} \alpha}
	+ \sum_{\alpha} (E_{\alpha} +J_{\alpha \alpha} )X_{\alpha \alpha}, \nonumber \\
	H_1 &= 
	 \sum_{\alpha \alpha'} J_{\alpha \alpha'} X_{\alpha \alpha'} ( -c_{\alpha} c_{\alpha'}^{\dag}),
\label{eq:hamil}
\end{align}
where $\epsilon_{\mib{k}}$ is the energy of conduction electrons with respect to the chemical potential, 
and $X_{\alpha \alpha'}$ is the $X$-operator on the local states $|\alpha\rangle$ defined by $X_{\alpha \alpha'}= |\alpha \rangle \langle \alpha' |$. 
$c_{\alpha} = N_0^{-1/2} \sum_{\mib{k}} c_{\mib{k} \alpha}$, with $N_0$ being number of sites, is the annihilation operator for the conduction electron at the impurity site. 
The order of the conduction operators in $H_1$ specifies the definition of the equal-time Green function used in the perturbation expansion. 
In order to distinguish the impurity contribution to the partition function, we factorize $Z$ as $Z=Z_f Z_{\rm c}$, where $Z_{\rm c}$ is the partition function without the impurity. 
The impurity contribution $Z_f$ is given by
\begin{align}
	Z_f =& \sum_{k=0}^{\infty} (-1)^k
	\int_0^{\beta} {\rm d}\tau_1 \cdots \int_{\tau_{k-1}}^{\beta} {\rm d}\tau_k
	\sum_{\alpha_1 \alpha'_1} \cdots \sum_{\alpha_k \alpha'_k} \nonumber \\
	&\times J_{\alpha_1 \alpha_1'} \cdots J_{\alpha_k \alpha_k'} \nonumber \\
	&\times
	s \prod_{\alpha} \langle T_{\tau} c_{\alpha}^{\dag}(\tau_1') c_{\alpha}(\tau_1'') \cdots
	 c_{\alpha}^{\dag}(\tau_{k_{\alpha}}') c_{\alpha}(\tau_{k_{\alpha}}'') \rangle_{\rm c} \nonumber \\
	&\times \text{Tr}_f \{ T_{\tau} {\rm e}^{-\beta H_f}
	 X_{\alpha_1 \alpha_1'}(\tau_1) \cdots X_{\alpha_k \alpha_k'}(\tau_k) \},
\end{align}
where conduction electron operators are grouped by component index, and averaged separately by $\langle \cdots \rangle_{\rm c} = Z_{\rm c}^{-1} \text{Tr}_{\rm c} \{ {\rm e}^{-\beta H_{\rm c}} \cdots \}$.
A resultant sign in the permutation is represented by $s$. 
$k_{\alpha}$ is the number of operators $c_{\alpha}^{\dag} c_{\alpha}$ for each component $\alpha$, and $\sum_{\alpha} k_{\alpha}=k$. 
Using Wick's theorem for the conduction electrons, we obtain 
\begin{align}
	Z_f =& \int {\rm D}[k] W_k, \nonumber \\
	W_k =& (-1)^k J_{\alpha_1 \alpha_1'} \cdots J_{\alpha_k \alpha_k'}
	\cdot s \prod_{\alpha} \det D_{\alpha}^{(k_{\alpha})} \nonumber \\
	&\times \text{Tr}_f \{ T_{\tau} {\rm e}^{-\beta H_f}
	 X_{\alpha_1 \alpha_1'}(\tau_1) \cdots X_{\alpha_k \alpha_k'}(\tau_k) \},
\label{eq:part_func}
\end{align}
where $\int {\rm D}[k]$ denotes the sum of $k$ and $\{ \alpha_i \}$ and the multiple integrals over $\{ \tau_i \}$. 
The $k_{\alpha}$ by $k_{\alpha}$ matrix $D_{\alpha}^{(k_{\alpha})}$ is defined by $(D_{\alpha}^{(k_{\alpha})})_{ij} = g_{\alpha} (\tau_i'' - \tau_j')$ with $g_{\alpha} (\tau) = -\langle T_{\tau} c_{\alpha}(\tau) c_{\alpha}^{\dag} \rangle_{\rm c}$. 

In the CT-QMC, statistical averages are evaluated by means of, for example, the Metropolis algorithm with the weight of Monte Carlo configurations given by eq.~(\ref{eq:part_func}). 
The implementation of the random walk will be discussed in \S3. 
During the simulations, it is enough to store only the inverse matrix $M_{\alpha}^{(k_{\alpha})}=(D_{\alpha}^{(k_{\alpha})})^{-1}$.
The matrix $M_{\alpha}^{(k_{\alpha})}$ is utilized to evaluate 
determinant ratios and is efficiently updated using fast-update formulae\cite{Rubtsov}.

\subsection{Impurity $t$-matrix}
The conduction electron Green function $G_{\alpha}(\tau, \tau'; \{ \tau_i \})$ for each configuration $\{ \tau_i \}$ is represented as follows:\cite{Rubtsov}
\begin{align}
	G_{\alpha}(\tau, \tau'; \{ \tau_i \}) &= g_{\alpha} (\tau- \tau') \nonumber \\
	 -&\sum_{ij} g_{\alpha} (\tau- \tau_j) (M_{\alpha}^{(k_{\alpha})})_{ji} g_{\alpha} (\tau_i - \tau').
\end{align}
This equation can be derived by using the fast-update formula for $M_{\alpha}^{(k_{\alpha})}$. 
An average over the Monte Carlo ensemble gives the physical Green function: $G(\tau, \tau') = \langle G(\tau, \tau'; \{ \tau_i \}) \rangle_{\rm MC}$. 
After the Fourier transform, we obtain
\begin{align}
	G_{\alpha}({\rm i}\epsilon_n) &= g_{\alpha} ({\rm i}\epsilon_n)
	 + g_{\alpha} ({\rm i}\epsilon_n) t_{\alpha}({\rm i}\epsilon_n) g_{\alpha} ({\rm i}\epsilon_n), \\
	t_{\alpha}({\rm i}\epsilon_n) &= -T\left< \sum_{ij} (M_{\alpha}^{(k_{\alpha})})_{ji}
	 {\rm e}^{{\rm i}\epsilon_n (\tau_j -\tau_i)} \right>_{\rm MC},
\end{align}
where $\epsilon_n=(2n+1)\pi T$ is the fermion Matsubara frequency. 
Since numerical summations over $i$ and $j$ are time consuming, it is more convenient to measure in imaginary time as follows:
\begin{align}
	t_{\alpha} (\tau) = -T \left< \sum_{ij} (M_{\alpha}^{(k_{\alpha})})_{ji} \delta(\tau, \tau_j - \tau_i) \right>_{\rm MC}.
\end{align}
In numerical calculations, the $\delta$-function is replaced by a rectangular function with a finite width, and $\tau$ is sampled in $\tau>0$ with use of the anti-periodicity as proposed in ref.~\citen{Werner}. 

We should note that the $t$-matrix in frequency space includes a constant term $t_{\alpha}^{(1)}$ in the first Born approximation
\begin{align}
	t_{\alpha}^{(1)} = J_{\alpha \alpha} \langle X_{\alpha \alpha} \rangle.
\label{eq:born_term}
\end{align}
Although $M_{\alpha}^{(k_{\alpha})}$ contains information of $t_{\alpha}^{(1)}$ at $\tau_j=\tau_i$, it is convenient to add it after the Fourier transformation rather than measure it in $\tau$-space. 
The constant is relevant for the calculation of the Kondo model as will be discussed in \S4. 
If the $t$-matrix is sampled in frequency space, the constant term is automatically included in $t_{\alpha}({\rm i}\epsilon_n)$.

\subsection{Two-particle correlation function}
We consider two-particle correlation functions $\chi_{\alpha \alpha'}(\tau)$ defined by $\chi_{\alpha \alpha'}(\tau) = \langle  T_{\tau} \tilde{X}_{\alpha \alpha}^{\rm H}(\tau) \tilde{X}_{\alpha' \alpha'}\rangle$, where the superscript H indicates the Heisenberg operator, and the tilde indicates deviation from the expectation value: $\tilde{X}_{\alpha \alpha} = X_{\alpha \alpha} - \langle X_{\alpha \alpha} \rangle$.
In the CT-QMC, $\langle T_{\tau} X_{\alpha \alpha}^{\rm H}(\tau') X_{\alpha' \alpha'}^{\rm H}(\tau'') \rangle$ is evaluated by
\begin{align}
	\left<
	\frac{ \text{Tr}_f \{ T_{\tau} {\rm e}^{-\beta H_f}
	 X_{\alpha_1 \alpha_1'}(\tau_1) \cdots X_{\alpha_k \alpha_k'}(\tau_k)
	  X_{\alpha \alpha}(\tau') X_{\alpha' \alpha'}(\tau'') \} }
	{ \text{Tr}_f \{ T_{\tau} {\rm e}^{-\beta H_f}
	 X_{\alpha_1 \alpha_1'}(\tau_1) \cdots X_{\alpha_k \alpha_k'}(\tau_k) \}}
	\right>_{\rm MC}.
\end{align}
In actual computations, it is not necessary to evaluate the matrix products for the trace. 
Instead, it is sufficient to judge whether $X_{\alpha \alpha}(\tau')$ and $X_{\alpha' \alpha'}(\tau'')$ are on segments of $\alpha$ and $\alpha'$ components respectively. In other words, we test whether $X_{\alpha \alpha}(\tau')$ and $X_{\alpha' \alpha'}(\tau'')$ are permitted by the $X$-operators in front and behind of them. 
By averaging over $\tau'$ with $\tau = \tau' - \tau''$, $\chi_{\alpha \alpha'}(\tau)$ is obtained with high accuracy. 
The equal-time correlation can be represented in terms of the mean occupation by $\chi_{\alpha \alpha'}(0) = \langle X_{\alpha \alpha} \rangle (\delta_{\alpha \alpha'} - \langle X_{\alpha' \alpha'} \rangle)$.

We also consider response functions for $M=\sum_{\alpha} m_{\alpha} X_{\alpha \alpha}$
\begin{align}
	\chi(\tau) = \langle T_{\tau} M^{\rm H}(\tau) M \rangle. 
\end{align}
By choosing proper $m_{\alpha}$ with $\sum_{\alpha} m_{\alpha}=0$, we can deal with magnetic, quadrupole and other moments within $N$ states. 
Provided the local levels are degenerate and the system has SU($N$) symmetry, the susceptibility is given by the Curie constant $C_N=N^{-1}\sum_{\alpha}m_{\alpha}^2$. 
The dynamical susceptibility $\chi(\tau)$ is then given in terms of $\chi_{\alpha \alpha'}(\tau)$ by
\begin{align}
	\frac{\chi(\tau)}{C_N}
	 = \sum_{\alpha}  \chi_{\alpha \alpha}(\tau)
	 - \frac{1}{N-1} \sum_{\alpha \neq \alpha'} \chi_{\alpha \alpha'} (\tau). 
\end{align}
The sums over $\alpha$ and $\alpha'$ improve the accuracy of sampling, although $\chi_{\alpha \alpha}$ and $\chi_{\alpha \alpha'}$ are 
identical for all $\alpha$ and for any combination of $\alpha \neq \alpha'$, respectively. 
The equal-time correlation is $\chi(0)/C_N = 1$. 
The static susceptibility is evaluated by integrating $\chi(\tau)$. 
Although the static susceptibility can also be obtained by measuring the expectation value of $M$ in the presence of a small external field, the evaluation using $\chi(\tau)$ gives results with higher precision.

\subsection{Specific heat}

Thermodynamic quantities can be evaluated from the single-particle Green function. 
The expression for the internal energy is obtained from the equation of motion for $G(\tau, \tau')$ in the limit $\tau' \rightarrow \tau +0$ \cite{Fetter-Walecka}.
A contribution $E_{\rm imp}$ of the impurity to the internal energy is given in terms of the impurity $t$-matrix $t_{\alpha}({\rm i}\epsilon_n)$ for forward scattering as follows:
\begin{align}
	E_{\rm imp} &= \langle H \rangle - \langle H_{\rm c} \rangle_{\rm c} \nonumber \\
	&= \sum_{\alpha} \Bigg[ E_{\alpha} \langle X_{\alpha \alpha} \rangle \nonumber \\
	&+ T\sum_n {\rm i}\epsilon_n
	 \left( \frac{1}{N_0} \sum_{\mib k} g_{{\mib{k}} \alpha}^2({\rm i}\epsilon_n) \right)
	t_{\alpha} ({\rm i}\epsilon_n) {\rm e}^{{\rm i}\epsilon_n \delta} \Bigg],
\label{eq:internal_energy}
\end{align}
where $\delta$ is a positive infinitesimal quantity. 

The specific heat $C$ is evaluated from the difference of $E_{\rm imp}$ at different temperatures. 
Assume that the internal energies at temperatures $T_0$ and $T_1$ ($T_0 < T_1$) are obtained as $E_0$ and $E_1$, respectively.
The specific heat at $(T_0 + T_1)/2$ is given by
\begin{align}
	C=\frac{E_1 - E_0}{T_1 - T_0},
\label{eq:heat}
\end{align}
and its statistical error $\Delta C$ is estimated by
\begin{align}
	\Delta C = \frac{\sqrt{(\Delta E_0)^2 + (\Delta E_1)^2}}{T_1 - T_0}, 
\end{align}
where $\Delta E_0$ and $\Delta E_1$ are standard deviations of $E_0$ and $E_1$, respectively. 
In this derivation, Gaussian distributions have been assumed. 
We note that the specific heat computed in eq.~(\ref{eq:heat}) includes, in addition to the statistical errors, an error due to the finite differences, which is proportional to powers of $T_1 - T_0$. 
Hence, in order to obtain a reasonable numerical accuracy, the statistical errors of the internal energy need to be smaller than $E_1-E_0$ when the temperature difference $T_1 - T_0$ is decreased.

\section{Monte Carlo Procedure}
In the CT-QMC method, we evaluate the statistical average of physical quantities by samplings with respect to the weight $W_k$ in eq.~(\ref{eq:part_func}).
The random walk in configuration space must satisfy the ergodicity and detailed balance conditions. 

Updates which change the order of $J$ are required for ergodicity, and updates which shift one of the operators increase sampling efficiency. 
In this section, we introduce two different algorithms for the random walk. 
The first, which changes the perturbation order by $\pm 1$ in each update, is most efficient. 
If some coupling constants are 0, however, the sampling may not be ergodic. 
For example, when the interaction lacks diagonal elements in the $N=2$ model, the 
perturbation order must be changed by $\pm 2$, 
because only even powers of $J$ contribute to the partition sum. 
In the general case, if only some coupling constants are finite, one needs to manipulate several operators (up to $N$) in one update.
The algorithm is presented in the latter of this section. 
Either algorithm should be chosen depends on the model.

\subsection{Method 1: manipulation of a segment}

\begin{figure}[t]
	\begin{center}
	\includegraphics[width=6cm]{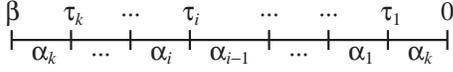}
	\end{center}
	\caption{A diagram representing the configuration of $\{ \tau_i \}$ and $\{ \alpha_i \}$ of order $J^k$.}
	\label{fig:diagram}
\end{figure}

\begin{figure}[t]
	\begin{center}
	\includegraphics[width=6cm]{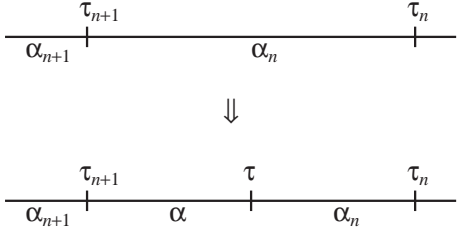}
	\end{center}
	\caption{Illustration of an insertion of a segment.}
	\label{fig:segment}
\end{figure}

A certain configuration of order $J^k$ is represented in terms of 
$\{ \tau_i \}=(\tau_1, \cdots, \tau_{k})$ and 
$\{ \alpha_i \}=(\alpha_1, \cdots, \alpha_{k})$.
These variables define the sequence of $X$-operators 
\begin{align}
	X_{\alpha_{k} \alpha_{k-1}}(\tau_{k}) \cdots X_{\alpha_i \alpha_{i-1}} (\tau_i) \cdots
	X_{\alpha_1 \alpha_{k}}(\tau_1).
\end{align}
The corresponding $c$-operators are given by
\begin{align}
	(-1)^{k+1}
	c_{\alpha_{k}}^{\dag} (\tau_1) c_{\alpha_{k}}(\tau_{k}) \cdots&
	c_{\alpha_{i}}^{\dag} (\tau_{i+1}) c_{\alpha_{i}}(\tau_{i}) \nonumber \\
	&\times \cdots
	c_{\alpha_{1}}^{\dag} (\tau_2) c_{\alpha_{1}}(\tau_{1}).
\end{align}
We represent the above configuration by a diagram as shown in Fig.~\ref{fig:diagram}. 
The most efficient update for $\{ \tau_i \}$ and $\{ \alpha_i \}$ is the addition or removal of a single element. 
We consider the process of adding $\tau$ and $\alpha$, which are randomly chosen in the range $[0, \beta)$ and from the $N$ components, respectively. 
If $\tau$ satisfies $\tau_{n+1} > \tau > \tau_{n}$, $\{ \tau_i \}$ and $\{ \alpha_i \}$ change into
$(\tau_1, \cdots, \tau_n, \tau, \tau_{n+1}, \cdots, \tau_{k})$ and 
$(\alpha_1, \cdots, \alpha_n, \alpha, \alpha_{n+1}, \cdots, \alpha_{k})$, respectively. 
Then one of the $X$-operators is altered as
\begin{align}
	X_{\alpha_{n+1} \alpha_n} (\tau_{n+1})
	\rightarrow
	X_{\alpha_{n+1} \alpha} (\tau_{n+1}) X_{\alpha \alpha_n} (\tau),
\end{align}
which corresponds to the change illustrated in Fig.~\ref{fig:segment}. 
Namely, the segment $\alpha$ is inserted between $\alpha_n$ and $\alpha_{n+1}$ with shortening of the segment $\alpha_n$. 
In the corresponding removal process, we erase one randomly chosen segment.

According to the detailed balance condition, the ratio of the transition probabilities should be
\begin{align}
	\frac{p( k \rightarrow k+1 )}{p( k+1 \rightarrow k )}
	= \frac{W_{k+1}}{W_k} \frac{N \beta}{k+1}. 
\end{align}
The factors $N$ and $\beta$ are due to the random choices of $\alpha$ and $\tau$, respectively, and $k+1$ due to that in the removal process. 
Since $W_{k}$ has the dimension of $J^k$, $W_{k+1} \beta/W_k$ is a dimensionless quantity. 
For $k \neq 0$, the ratio $W_{k+1}/W_k$ is given by
\begin{align}
	\frac{W_{k+1}}{W_k}
	=& \frac{J_{\alpha_{n+1} \alpha} J_{\alpha \alpha_n}}{J_{\alpha_{n+1} \alpha_n}}
	\exp[ -l(E_{\alpha} - E_{\alpha_n}) ] \nonumber \\
	&\times \frac{\det D_{\alpha}^{(+)}}{\det D_{\alpha}}
	\frac{\det \tilde{D}_{\alpha_n}}{\det D_{\alpha_n}},
\label{eq:prob_seg}
\end{align}
where $l=\tau_{n+1} - \tau$ is the length of the new segment. 
$D_{\alpha}^{(+)}$ is the matrix with $c_{\alpha}^{\dag}(\tau_{n+1}) c_{\alpha}(\tau)$ added to $D_{\alpha}$, and $\tilde{D}_{\alpha_n}$ is the matrix with one of the operators shifted in time according to $c_{\alpha_n}^{\dag}(\tau_{n+1}) \rightarrow c_{\alpha_n}^{\dag}(\tau)$. 
The ratio of determinants can be evaluated using fast-update formulae\cite{Rubtsov}. 
If $\alpha = \alpha_n$ in Fig.~\ref{fig:segment}, the change is just an addition of a diagonal element $X_{\alpha \alpha}(\tau)$, so that eq.~(\ref{eq:prob_seg}) is reduced to
\begin{align}
	\frac{W_{k+1}}{W_k}
	= -J_{\alpha \alpha}
	\frac{\det D_{\alpha}^{(+)}}{\det D_{\alpha}}.
\end{align}
Here $D_{\alpha}^{(+)}$ is a matrix in which $c_{\alpha}^{\dag}(\tau) c_{\alpha}(\tau+0)$ is added to the original one. 
The equal-time Green function in $D_{\alpha}^{(+)}$ should be $g_{\alpha} (+0)$ to keep the probability positive. 
This is verified by taking the CS limit in the corresponding formula of the Anderson model (see Appendix A).
In the case of $k=0$, we should average over the local states since all states contribute to the trace. 
Then $W_1/W_0$ is given by
\begin{align}
	\frac{W_1}{W_0}
	= -J_{\alpha \alpha} \rho_{\alpha} g_{\alpha}(+0),
\label{eq:prob_seg0}
\end{align}
where
$\rho_{\alpha}$ is defined by $\rho_{\alpha} = \exp( -\beta E_{\alpha}) / \sum_{\alpha'} \exp( -\beta E_{\alpha'})$.

The ratios of the weights in eqs.~(\ref{eq:prob_seg})--(\ref{eq:prob_seg0}) change their signs depending on the signs of the coupling constants. 
We have confirmed by numerical calculations that the probability remains positive in the case of antiferro-couplings, i.e., $J_{\alpha \alpha'}>0$. 
This is consistent with the fact that the CS model with antiferro-couplings is derived from the Anderson model, where the minus sign problem does not appear. 
We also note that we could consider another process in which $\alpha$ is inserted between $\alpha_{n-1}$ and $\alpha_n$. 
It corresponds to a diagram where $\alpha$ and $\alpha_n$ are exchanged in Fig.~\ref{fig:segment}. 
However, the ergodicity condition can be satisfied without this process.

\subsection{Method 2: manipulation of a set of operators}

\begin{figure}[t]
	\begin{center}
	\includegraphics[width=6cm]{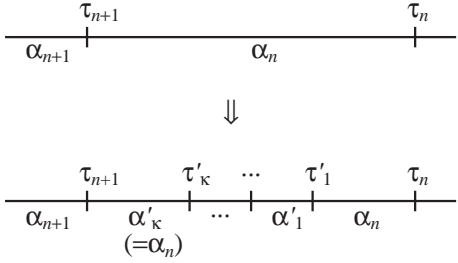}
	\end{center}
	\caption{Illustration of an addition of an irreducible operator.}
	\label{fig:irr_op}
\end{figure}

If some coupling constants are zero, the ergodicity may require updates that change 
the diagrams by several perturbation orders. 
At most $N$ operators should be inserted and removed in one update. 
To this end, we define irreducible sets of operators, which consist of $\kappa$ operators with a set of $\kappa$ distinct components $(\alpha'_1, \cdots, \alpha'_{\kappa})$.
Then the irreducible sets of $X$-operators are given by
\begin{align}
	X_{\alpha'_{\kappa} \alpha'_{\kappa-1}}(\tau'_{\kappa}) \cdots
	X_{\alpha'_j \alpha'_{j-1}}(\tau'_j) \cdots
	X_{\alpha'_1 \alpha'_{\kappa}}(\tau'_1).
\end{align}
We consider a process of insertion and removal of the irreducible operator, which is illustrated in Fig.~\ref{fig:irr_op}. 

To determine an irreducible operator, we begin by randomly choosing $\kappa$ from 1 to $N$ and $\tau'_1$ from the range $[0, \beta)$.
If $\tau'_1$ satisfies $\tau_{n+1} > \tau'_1 > \tau_n$, $\alpha'_{\kappa}$ is fixed as $\alpha'_{\kappa}=\alpha_n$. 
The remaining components $\alpha'_j$ ($1\leq j < \kappa$) are determined so as to be different from each other. 
The imaginary times $\tau'_j$ ($1<j \leq \kappa$) are randomly chosen from the restricted range $(\tau'_{j-1}, \tau_{n+1})$. 
In the case of a removal process, we choose one, say $\alpha_n$, from all the segments. 
If two identical components appear 
between the selected segment and the next segment with component $\alpha_n$, the operator set is not irreducible and hence the update is rejected at this stage. 
If this is not the case, we proceed to attempt a removal of the operators.

From the detailed balance condition, we obtain the update probability of the above process for $k \neq 0$ as
\begin{align}
	\frac{p( k \rightarrow k+\kappa )}{p( k+\kappa \rightarrow k )}
	= \frac{W_{k+\kappa}}{W_k} \frac{N!}{(N-\kappa)!} \frac{1}{k+\kappa} \prod_{j=1}^{\kappa} l^{\rm (max)}_j, 
\end{align}
where $l_j^{\rm (max)}$ is given by $l_1^{\rm (max)}=\beta$ and $l_j^{\rm (max)}=\tau_{n+1} - \tau'_{j-1}$ for $j > 1$. 
The factor $N! / (N-\kappa)!$ is the inverse of the probability that the corresponding irreducible operator is chosen. 
The ratio $W_{k+\kappa}/W_k$ is given by
\begin{align}
	\frac{W_{k+\kappa}}{W_k}
	&= -\prod_{j=1}^{\kappa} \left\{ J_{\alpha'_{j} \alpha'_{j-1}}
	 \exp [ - l_j ( E_{\alpha'_j} -E_{\alpha_n} ) ]
	 \frac{\det D_{\alpha'_j}^{(+)}}{\det D_{\alpha'_j}} \right\},
\label{eq:prob_irr_op}
\end{align}
where $l_j$ is a length of the segment $\alpha'_j$. 
$D_{\alpha'_j}^{(+)}$ is a matrix 
to which $c_{\alpha'_j}^{\dag} (\tau'_{j+1}) c_{\alpha'_j} (\tau'_j)$ has been added, with $\tau'_{\kappa+1} = \tau'_1$. 
In the case of $k=0$, the above procedure is applicable with $\tau_n=0$ and $\tau_{n+1}=\beta$. 
One then has to choose $\alpha'_{\kappa}$ at random, and take the trace over the local states. 
The transition probability for $k=0$ becomes
\begin{align}
	\frac{p(0 \rightarrow \kappa)}{p(\kappa \rightarrow 0)}
	&= \frac{W_{\kappa}}{W_0} \frac{N \cdot N!}{(N-\kappa)!} \prod_{j=1}^{\kappa} l^{\rm (max)}_j. 
\end{align}
The additional factor $N$ originates in the random choice of $\alpha'_{\kappa}$. 
We note that the factor $1/(k+\kappa)$ in eq.~(\ref{eq:prob_irr_op}) is not necessary for $k=0$ because no choice is made in the removal process. 
The ratio $W_{\kappa}/W_0$ is given by
\begin{align}
	\frac{W_{\kappa}}{W_0}
	&= - Z_{f0}^{-1}
	\prod_{j=1}^{\kappa} \left\{ J_{\alpha'_{j} \alpha'_{j-1}}
	 \exp ( - l_j  E_{\alpha'_j} )
	 \det D_{\alpha'_j}^{(+)} \right\}, 
\end{align}
where we define $Z_{f0}$ by $Z_{f0}=\sum_{\alpha} \exp(-\beta E_{\alpha})$.

\section{The Kondo Model}
When we deal with $S=1/2$ systems without orbital degeneracy, the Kondo model is more favorable than the CS model, because of the particle-hole symmetry in the former. 
The algorithm for the CS model is applicable to the Kondo model after some modifications. 

The Kondo model is given by
\begin{align}
	H = \sum_{\mib{k} \sigma} \epsilon_{\mib{k}} c_{\mib{k} \sigma}^{\dag} c_{\mib{k} \sigma}
	+ J \mib{S} \cdot \mib{\sigma}_c,
\label{eq:Kondo}
\end{align}
where $\mib{\sigma}_c$ is the Pauli matrix for conduction electrons. 
The Hamiltonian is rewritten in terms of the CS interaction as follows:
\begin{align}
	H = \sum_{\mib{k} \sigma} \epsilon_{\mib{k}} c_{\mib{k} \sigma}^{\dag} c_{\mib{k} \sigma}
	 + v \sum_{\sigma} c_{\sigma}^{\dag} c_{\sigma}
	 + J \sum_{\sigma \sigma'} f_{\sigma}^{\dag} f_{\sigma'} c_{\sigma'}^{\dag} c_{\sigma},
\label{eq:Kondo_CS}
\end{align}
where $v=-J/2$. 
Therefore the algorithm for the CS model is applicable by including the potential scattering term into $H_0$. 

We introduce the Green function including the potential scattering, $\tilde{g}(z)$, which is a scalar in the spin indices. 
In terms of the bare Green function $g(z)$, $\tilde{g}(z)$ is expressed as
\begin{align}
	\tilde{g} = \frac{g}{1 - v g}.
\end{align}
In the simulation of the CS model, $g(z)$ is replaced by $\tilde{g}(z)$. 
Within the CT-QMC, the impurity $t$-matrix $t_J(z)$ is computed with respect to $\tilde{g}(z)$.
To obtain the $t$-matrix $t(z)$ of the Kondo model, eq.~(\ref{eq:Kondo}), the contribution of the potential scattering should be subtracted from $\tilde{g}(z)$. 
The full Green function $G(z)$ can be expressed as
\begin{align}
	G = \tilde{g} + \tilde{g} t_J \tilde{g}
	 = g + g t g. 
\label{eq:Kondo_G}
\end{align}
Solving eq.~(\ref{eq:Kondo_G}) with respect to $t(z)$, we obtain
\begin{align}
	t = \frac{v}{1-vg} + \frac{t_J}{(1-vg)^2},
\label{eq:tmat_kondo}
\end{align}
where the first term is the $t$-matrix due to the potential scattering.
Concerning the two-particle correlation function, no modification is required, because it does not depend on the selection of the basis set in the perturbation expansion.

\section{Imaginary-Time Data and Static Quantities}

We apply our algorithm to a model with SU($N$) symmetry. 
In this case, both types of updates introduced in \S3 are available. 
We have confirmed that the results of the two methods agree within error bars. 
Therefore we use 
the more efficient rank one updates for the rest of the calculations. 

We use a constant density of states for the conduction electrons
\begin{align}
	-\frac{1}{\pi} \text{Im} g_{\alpha} (\epsilon + {\rm i}\delta) = \rho_0 \ \theta(D-|\epsilon|),
\label{eq:dos_c}
\end{align}
where $\theta(\epsilon)$ is a step function and $\rho_0=1/2D$. 
In the Matsubara formalism, the Green function is represented as $g({\rm i}\epsilon_n) = -2{\rm i}\rho_0 \arctan(D/\epsilon_n)$ and $N_0^{-1} \sum_{\mib{k}} g_{\mib{k}}^2({\rm i}\epsilon_n) = -(\epsilon_n^2 + D^2)^{-1}$.
We choose the unit $D=1$ in this paper. 
The standard deviations in the MC ensembles are evaluated from 20 averages. 
In MC simulations, we have observed minus sign probabilities at temperatures lower than the Kondo temperature. 
However, they appear only at the rate of about one in $10^7$ updates, and 
may be due to rounding errors. 

In the following, we show numerical results for physical quantities without analytic continuation, such as the static susceptibility and the specific heat. 
These results are, within error bars, exact. 
Spectral function in real frequencies will be shown in the next section.

\subsection{$N=1$: potential scattering}

The Hamiltonian, eq.~(\ref{eq:hamil}), is trivially solvable when $N=1$, which corresponds to potential scattering. 
The exact $t$-matrix for the potential scattering is given by the first term on the right-hand side of eq.~(\ref{eq:tmat_kondo}) with $v=J$, which 
takes into account an infinite sequence of scattering events. 
We apply our algorithm to the potential scattering to compare with the exact solution, prior to applications to $N>1$. 
\begin{figure}[tb]
	\begin{center}
	\includegraphics[width=8cm]{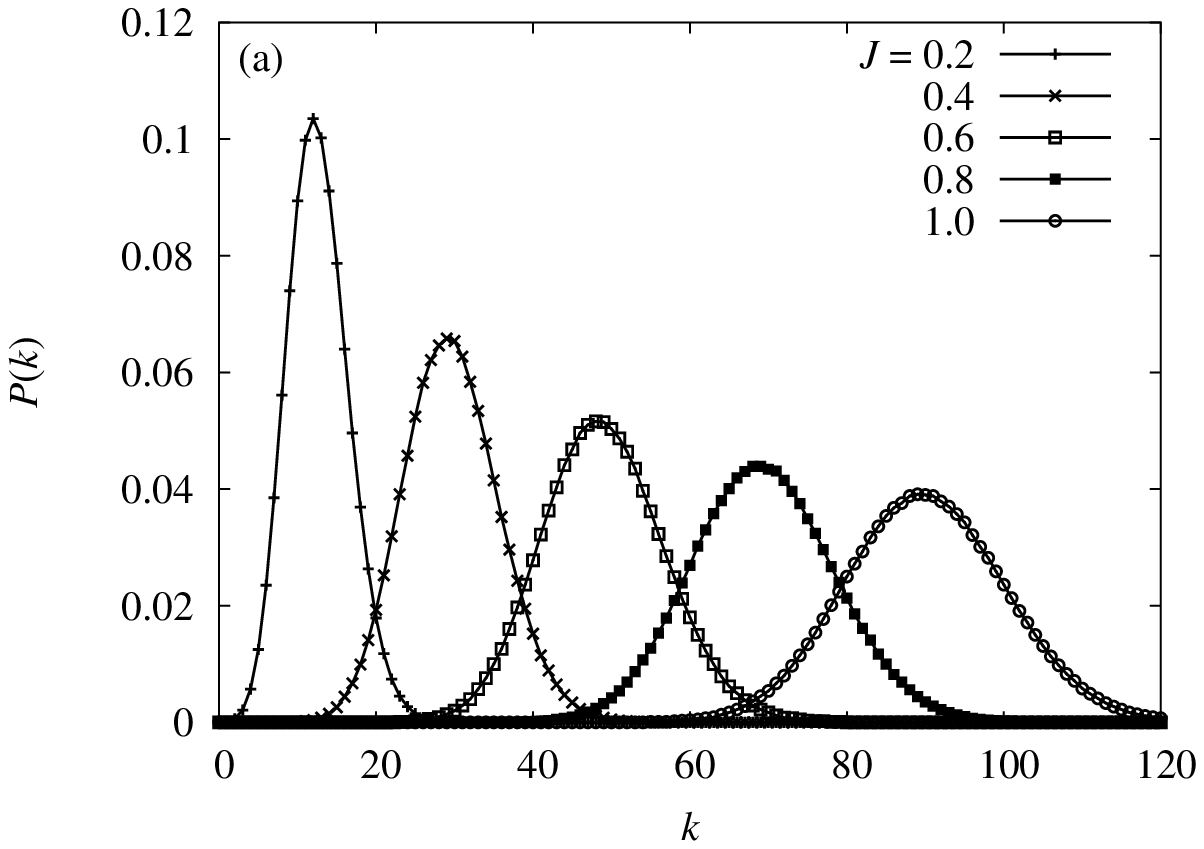}
	\includegraphics[width=8cm]{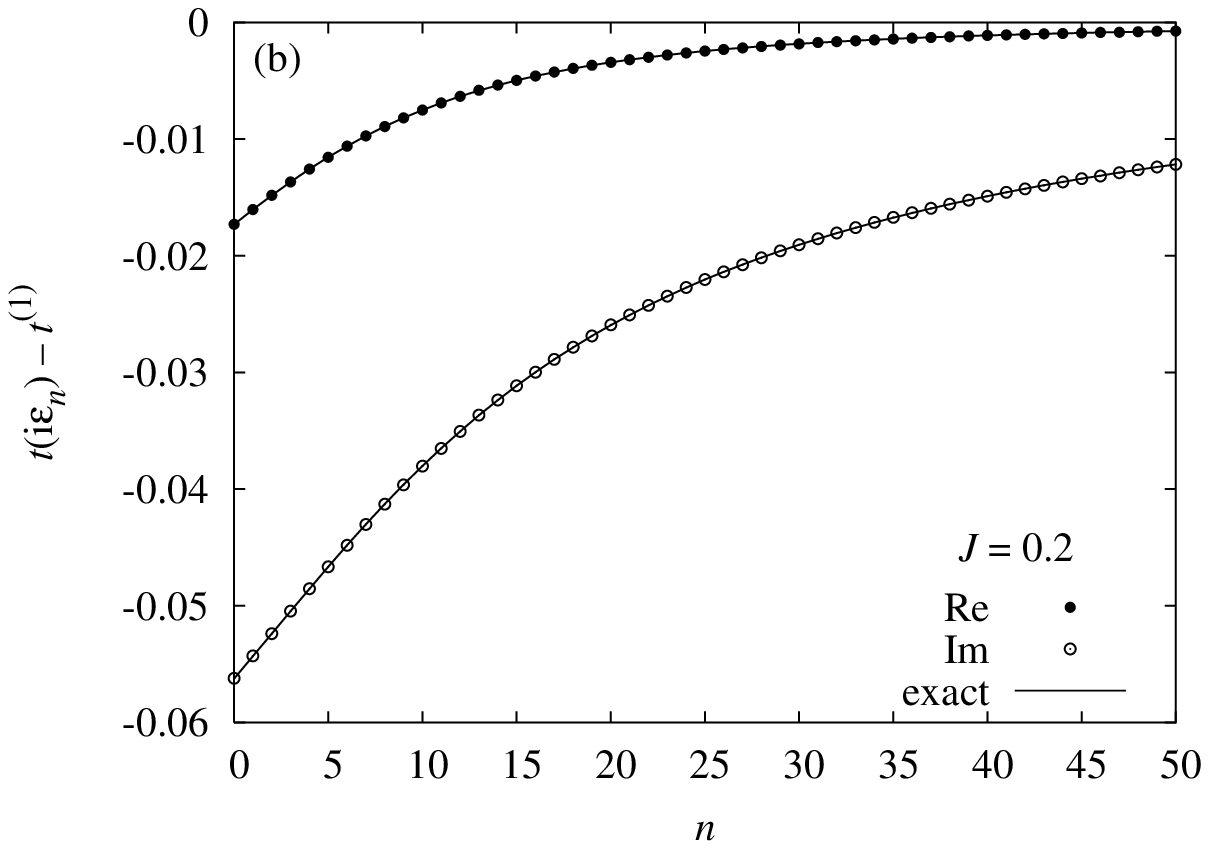}
	\end{center}
	\caption{(a) Probability distribution $P(k)$ for terms of order $J^k$, and (b) $t$-matrix $t({\rm i}\epsilon_n)$ for $N=1$. The temperature is chosen as $T=0.01$. }
	\label{fig:pot}
\end{figure}
Figure~\ref{fig:pot}(a) shows the probability of appearance of perturbation terms of order $J^k$ in the Monte Carlo simulations. 
Although the exact results include infinite series of scattering, only terms of finite order of $J$ are significant in Monte Carlo simulations. 
The center of the distribution increases as $J$ increases. 
Results for the $t$-matrix in the Monte Carlo ensemble average are shown in Fig.~\ref{fig:pot}(b) together with the analytical result. 
The term $t^{(1)}$ given by eq.~(\ref{eq:born_term}) is subtracted in this figure. 
The exact results are completely reproduced,  
which demonstrates the convergence of the perturbation series, and shows that sampling finite perturbation orders is sufficient.

\subsection{$N=8$: large-$N$ case}
The present algorithm can handle arbitrary $N$. 
We first take $N=8$ and give exemplary results of static quantities as well as raw data in the imaginary-time domain. 
\begin{figure}[tb]
	\begin{center}
	\includegraphics[width=8cm]{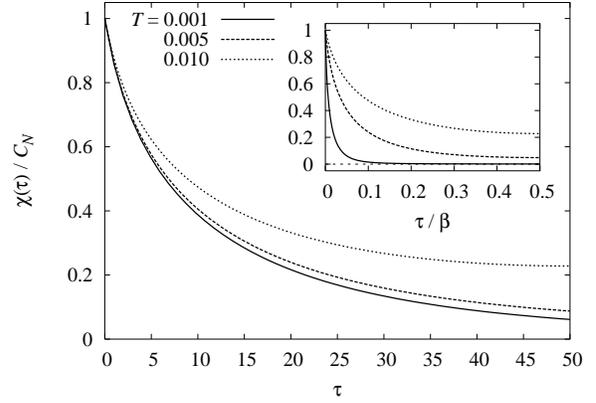}
	\end{center}
	\caption{Dynamical susceptibility $\chi(\tau)$ in the imaginary-time domain for $N=8$ and $J=0.075$. }
	\label{fig:n8-chi_tau}
\end{figure}
Figure~\ref{fig:n8-chi_tau} shows the dynamical susceptibility $\chi(\tau)$ for several values of $T$. 
We have plotted with lines because the intervals between data points are fine enough, and the associated errors are negligible. 
It turns out that the reduction of the correlation, which starts at $\chi(\tau=0)=1$, becomes more rapid for lower temperatures.
This implies 
Kondo screening at low temperatures. 
The inset shows $\chi(\tau)$ against $\tau/\beta$. 
At $T=0.001$, the correlation almost disappears at $\tau=\beta/2$, while correlations remain for $T=0.005$. 
This indicates complete screening of the local moment at $T=0.001$. 

\begin{figure}[tb]
	\begin{center}
	\includegraphics[width=8cm]{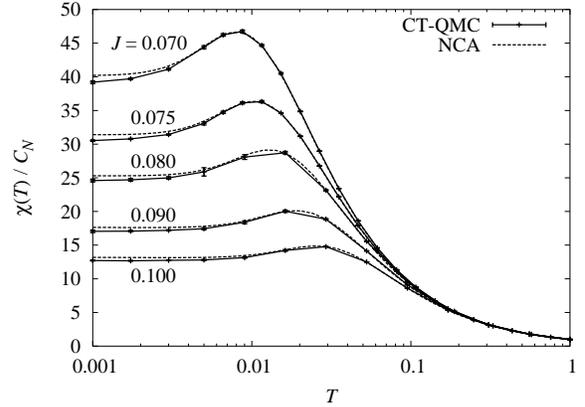}
	\end{center}
	\caption{Temperature dependence of the static susceptibility $\chi$ for $N=8$ and several values of $J$. 
	Dashed lines are results computed in the NCA. }
	\label{fig:n8-suscep}
\end{figure}
The static susceptibility $\chi(\omega=0)$ can be evaluated by integrating $\chi(\tau)$.
Figure~\ref{fig:n8-suscep} shows the temperature dependence of the static susceptibility for several values of $J$ for $N=8$. 
For comparison, we plot results computed in the non-crossing approximation (NCA), which incorporates terms up to the next-leading order in the $1/N$ expansion by integral equations\cite{nca1, Bickers}. 
The CT-QMC results are in excellent agreement with the NCA at high temperatures. 
On the other hand, deviations are visible at lower temperatures. 
This is due to the inaccuracies of the NCA at low temperatures and low frequencies\cite{nca3}. 
We have confirmed in fact that the incorrect upturn appears in the NCA results for smaller $N$, where the accuracy of the $1/N$ expansion diminishes. 
On the other hand, the CT-QMC produces proper values of the static susceptibility at all temperatures.

\begin{figure}[tb]
	\begin{center}
	\includegraphics[width=8cm]{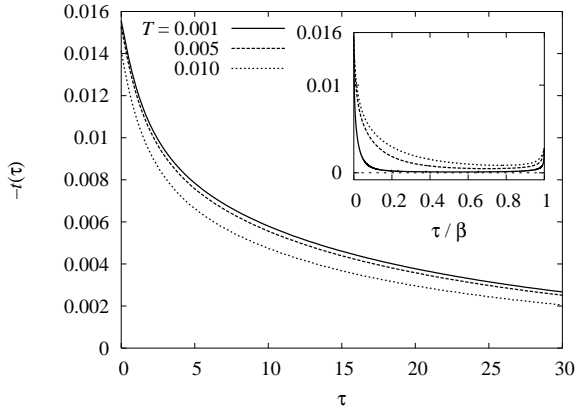}
	\end{center}
	\caption{The impurity $t$-matrix $t_{\alpha}(\tau)$ in the imaginary-time domain for $N=8$ and $J=0.075$. }
	\label{fig:n8-t_matrix}
\end{figure}
We next show results for the impurity $t$-matrix. 
Figure~\ref{fig:n8-t_matrix} shows the $t$-matrix $t(\tau)$ in the imaginary-time domain for the same parameters as in Fig.~\ref{fig:n8-chi_tau}. 
Statistical errors are so small that we have plotted with lines as in the case of $\chi(\tau)$. 
As temperature decreases, $-t(\tau)$ increases. 
Correspondingly, the discontinuity at the boundary $t(-0)-t(+0)$ increases as temperature decreases. 
Since the discontinuity is given by $-\pi^{-1} \int \text{Im} t(\omega) {\rm d}\omega$, the increase indicates enhanced scattering due to the impurity.
This corresponds to the formation of the Kondo singlet.

\begin{figure}[tb]
	\begin{center}
	\includegraphics[width=8cm]{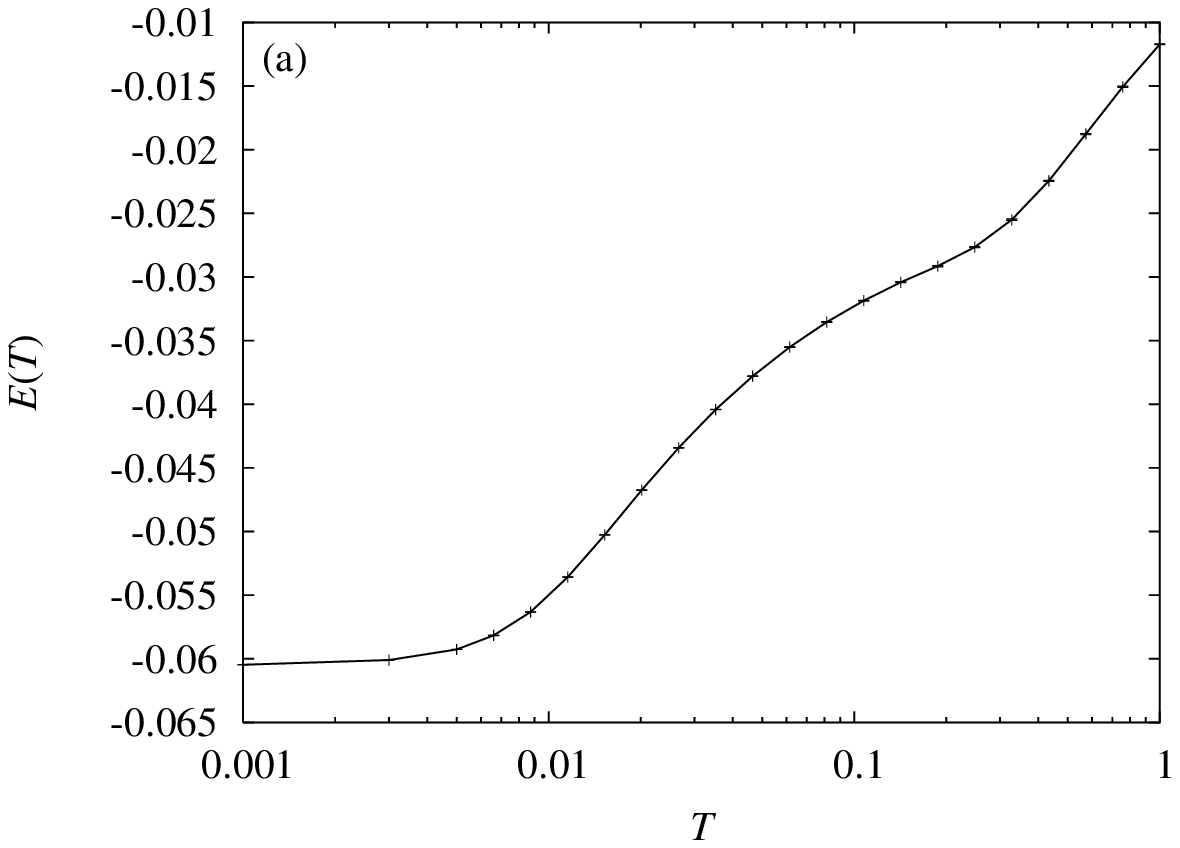}
	\includegraphics[width=8cm]{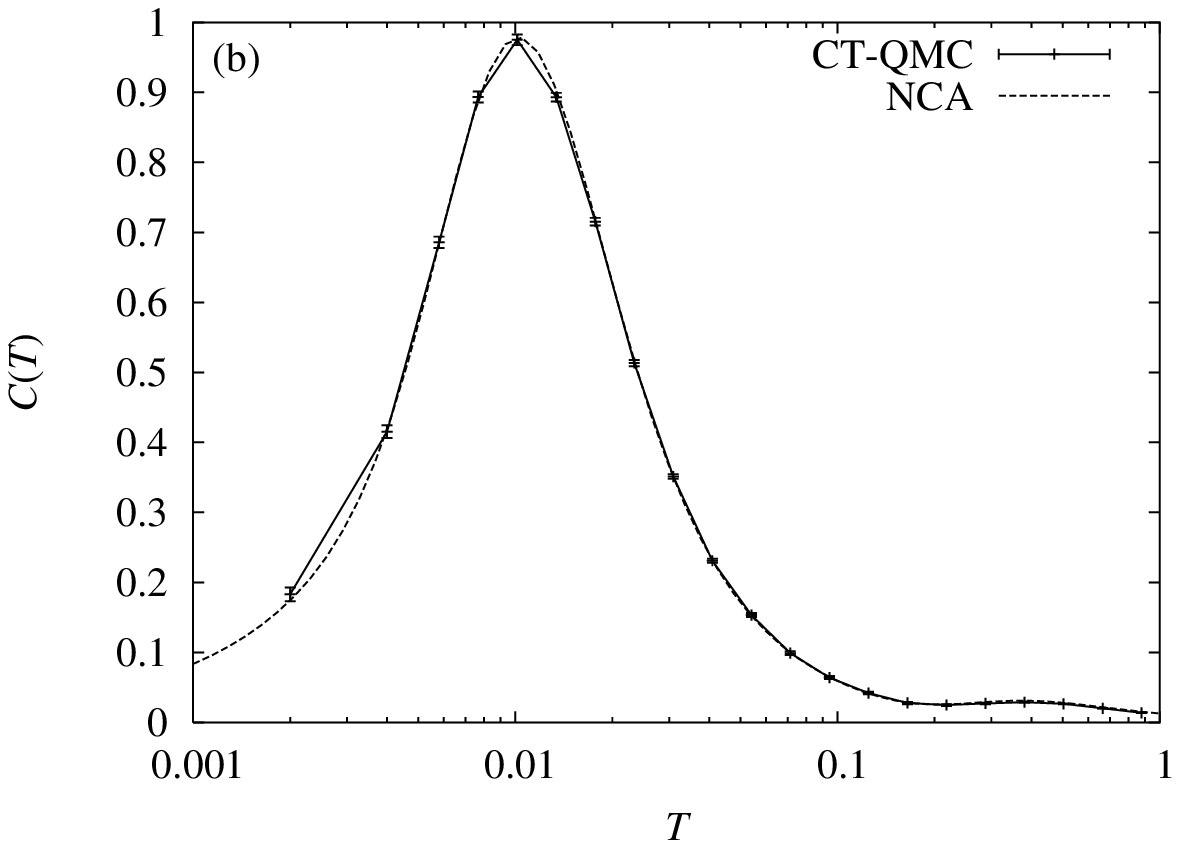}
	\end{center}
	\caption{Temperature dependence of the internal energy $E(T)$ and the specific heat $C(T)$ for $N=8$ and $J=0.075$. 
	The dashed line is the result computed in the NCA. }
	\label{fig:n8-heat}
\end{figure}
The impurity $t$-matrix with imaginary frequency describes thermodynamic quantities as shown in eq.~(\ref{eq:internal_energy}). 
The temperature dependence of the internal energy $E(T)$ is shown in Fig.~\ref{fig:n8-heat}(a). 
The statistical errors are invisible in this figure. 
The data give the specific heat $C(T)$ via eq.~(\ref{eq:heat}). 
Figure~\ref{fig:n8-heat}(b) shows the temperature dependence of $C(T)$ plotted together with results from the NCA\cite{Otsuki_thermo}. 
Statistical errors become larger at low temperatures due to the small mesh of temperatures.
The CT-QMC agrees with the NCA within error bars, and reproduces a peak due to the Kondo effect.

\subsection{Comparison between different $N$}
So far we have presented imaginary-time data and resultant static quantities for $N=8$. 
Next we examine different values of $N$. 
Coupling constants are chosen so as to fix $N J$ to yield almost equal values of the Kondo temperature. 
We define the Kondo temperature $T_{\rm K}$ by means of the static susceptibility $\chi$ by $T_{\rm K} = C_N / \chi$ at low enough temperature. 
Table~\ref{tab:param} shows $\chi$ at $T=0.001$ and calculated values of $T_{\rm K}$ for $N=2$, 4, 6 and 8. 
For all $N$, $T_{\rm K}$ is less than 1/30 of $D$, and therefore can be considered small compared to $D$. 
\begin{table}[tb]
	\begin{center}
	\caption{The static susceptibility $\chi$ and the Kondo temperatures $T_{\rm K}$ estimated by $T_{\rm K}=C_N / \chi$ at $T=0.001$. Values in bracket are the standard deviations.\\}
	\label{tab:param}
	\begin{tabular}{cccc}
	\hline
	$N$ & $J$ & $\chi(T=0.001) / C_N$ & $T_{\rm K}$ \\
	\hline
	2 & 0.300 & 30.95 (0.24) & 0.0323 (0.0002) \\
	4 & 0.150 & 31.67 (0.14) & 0.0316 (0.0001) \\
	6 & 0.100 & 31.03 (0.11) & 0.0322 (0.0001) \\
	8 & 0.075 & 30.53 (0.10) & 0.0328 (0.0001) \\
	\hline
	\end{tabular}
	\end{center}
\end{table}
\begin{figure}[tb]
	\begin{center}
	\includegraphics[width=8cm]{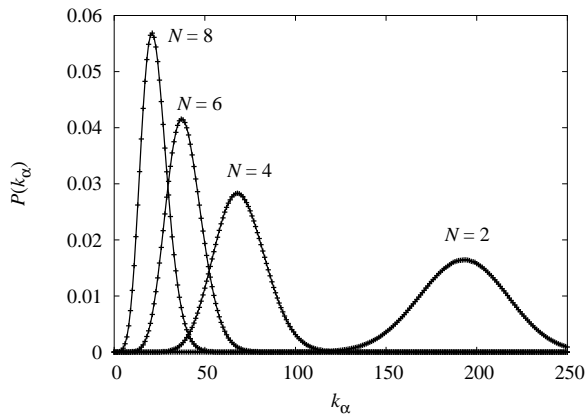}
	\end{center}
	\caption{Probability distributions $P(k_{\alpha})$ at $T=0.001$. Parameters and $T_{\rm K}$ are listed in Table~\ref{tab:param}. }
	\label{fig:n2468-stat}
\end{figure}
Figure~\ref{fig:n2468-stat} shows probabilities $P(k_{\alpha})$ of appearance of $k_{\alpha}$ for different $N$ at $T=0.001$. 
The size of the matrix $D_{\alpha}$ is equal to $k_{\alpha}$. 
It turns out that the peak of $P(k_{\alpha})$ shifts to higher values of $k_{\alpha}$ for smaller $N$. 
This is because the power of $J$ is divided into $N$ components, resulting in the decrease of $k_{\alpha}$. 
Consequently, a larger value of $N$ reduces the computational burden, and makes it possible to reach temperatures much lower than $T_{\rm K}$.

\begin{figure}[tb]
	\begin{center}
	\includegraphics[width=8cm]{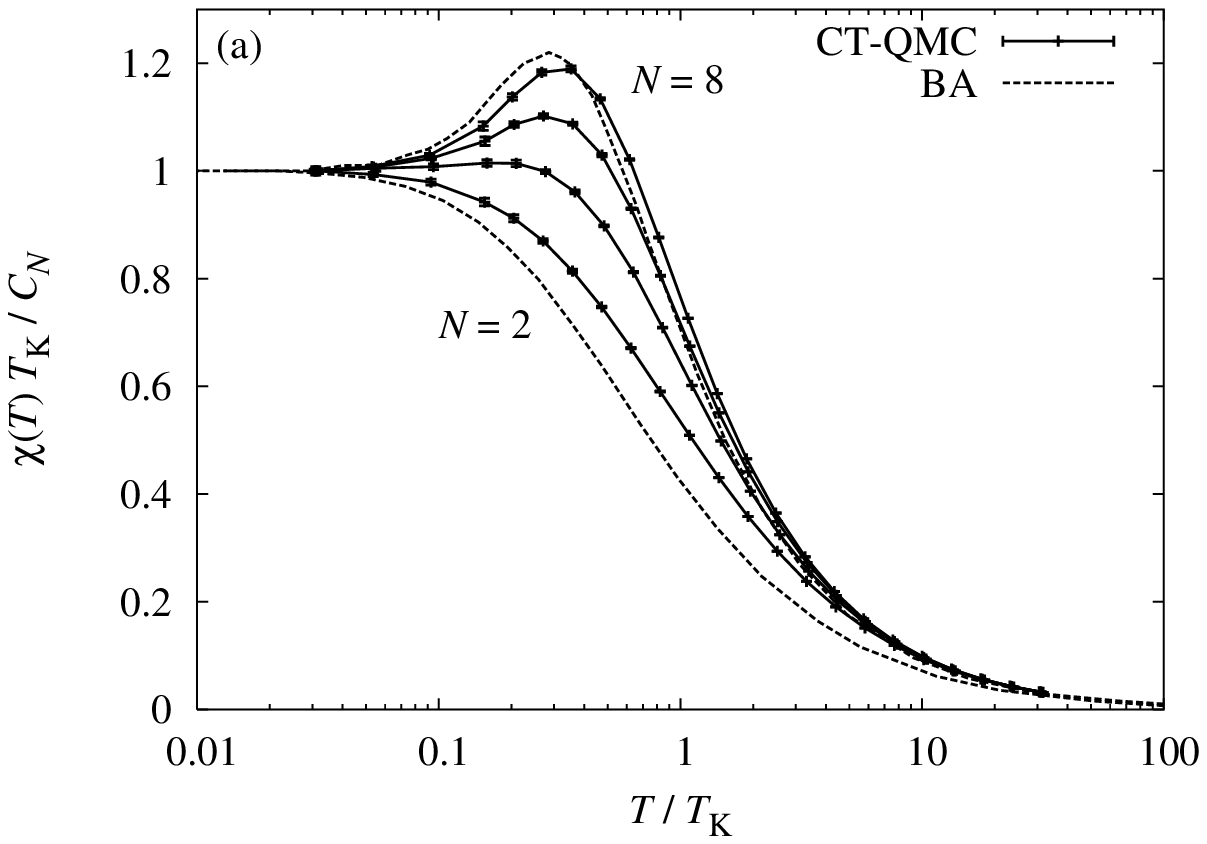}
	\includegraphics[width=8cm]{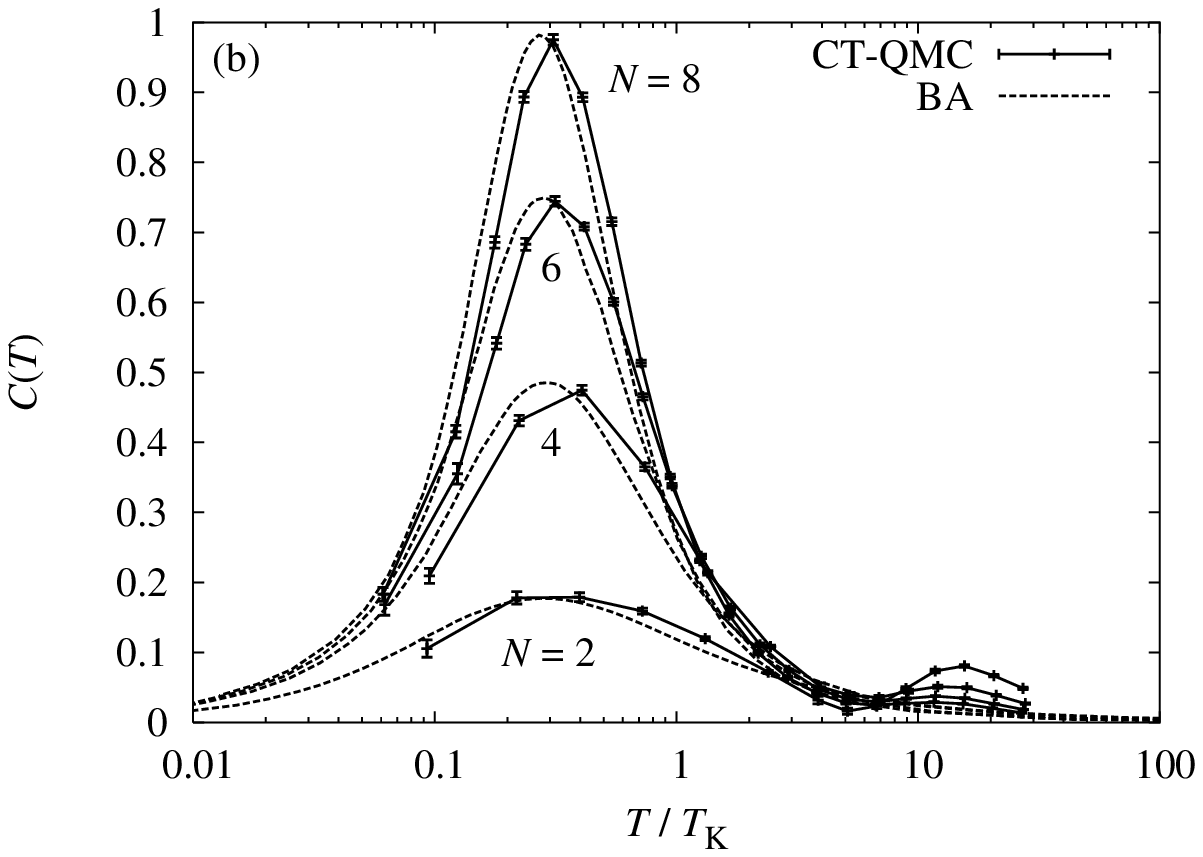}
	\end{center}
	\caption{Temperature dependence of (a) the static susceptibility and (b) the specific heat for $N=2$, 4, 6 and 8. 
	Parameters and $T_{\rm K}$ are listed in Table~\ref{tab:param}. Dashed lines are the Bethe ansatz solution\cite{Rajan}. }
	\label{fig:n2468-suscep}
\end{figure}
For static quantities, exact solutions have been obtained based on the Bethe ansatz\cite{Rajan}. 
Temperature dependences of all physical quantities are given through a single energy scale. 
Hence we can compare our results with the exact solutions, using the $T_{\rm K}$ determined above. 
We note that the characteristic energy $T_0$ in ref.~\citen{Rajan} relates to our Kondo temperature $T_{\rm K}$ by $T_{\rm K}=2\pi T_0 / N$. 
Figure~\ref{fig:n2468-suscep}(a) shows the static susceptibility as a function of $T/T_{\rm K}$ for the parameters listed in Table.~\ref{tab:param}. 
Dashed lines are the Bethe ansatz solutions for $N=2$ and 8. 
Comparison between our results and the exact solution reveals that $T_{\rm K}$ determined by $T_{\rm K}=C_N / \chi$ systematically deviates from the exact results. 
This is ascribed to the finite cutoff of the conduction band in our model, which is shown to enhance the static susceptibility. 
Consequently, $T_{\rm K}$ determined by $\chi$ tends to be smaller than the Bethe ansatz value. 
The effect of the finite band width will be discussed in detail later. 
Temperature dependences of the specific heat are shown in Fig.~\ref{fig:n2468-suscep}(b). 
Large peaks around $T/T_{\rm K} \sim 0.3$ are due to the Kondo effect, while small peaks at about $T/T_{\rm K} \sim 10$ originate from the cutoff of the conduction band. 
We recognize systematic deviations between our results and the Bethe ansatz solutions, corresponding to those found in the static susceptibility. 
However, the overall behavior agrees well if we scale temperature independently of the static susceptibility.

\subsection{The Kondo model}
We can deal with the Kondo model in the way presented in \S4. 
In applying our algorithm for the CS model to the Kondo model, we use unperturbed Green functions without particle-hole symmetry, although the original model is particle-hole symmetric. 
Therefore it is a strict check of our algorithm whether the particle-hole symmetry is recovered after solving the CS interactions. 
\begin{figure}[tb]
	\begin{center}
	\includegraphics[width=8cm]{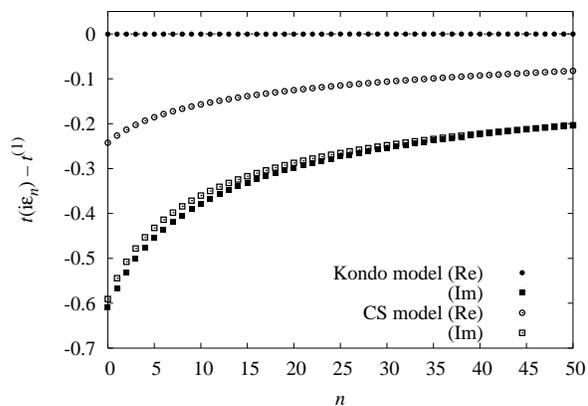}
	\end{center}
	\caption{$t$-matrix $t({\rm i}\epsilon_n)$ of the Kondo model and the CS model for $J=0.3$ and $T=0.001$.}
	\label{fig:kondo-t_omega}
\end{figure}
Figure~\ref{fig:kondo-t_omega} shows the $t$-matrix $t({\rm i}\epsilon_n)$ for $J=0.3$ and $T=0.001$. 
For comparison, a result for the CS model is plotted for the same coupling constant and temperature. 
In the imaginary-time representations, $\text{Re } t_{\alpha}({\rm i}\epsilon_n)=0$ indicates symmetry with respect to particle-hole excitations. 
It is verified that the result for the Kondo model keeps the particle-hole symmetry, while the CS model does not due to the potential scattering.

\section{Dynamical Quantities with Analytic Continuation}
We proceed to spectral functions in real frequencies. 
To perform analytic continuations, we employ two kinds of conventional methods:
 the Pad\'e approximation\cite{pade} and the maximum entropy method (MEM)\cite{Jarrell-Gubernatis}. 
The Pad\'e approximation fits data at the Matsubara frequencies in the upper-half plane with use of a rational function, and extrapolates onto the real axis. 
Since the Fourier transform drives away noises in imaginary-time data to high frequencies, reasonable accuracy is expected at low frequencies. 
However the extrapolation is sensitive to statistical and numerical errors, especially at high energies which are far from the imaginary axis. 
If the data include significant errors, the Pad\'e approximation does not work in general. 
In this case, the statistical errors can be taken into account by the MEM. 
The MEM derives spectral functions from data in the imaginary-time representation without Fourier transformation.
In this subsection, we shall show results by the Pad\'e approximation, since our data have been obtained with sufficient accuracy.

\subsection{Impurity $t$-matrix}

\begin{figure}[tb]
	\begin{center}
	\includegraphics[width=7cm]{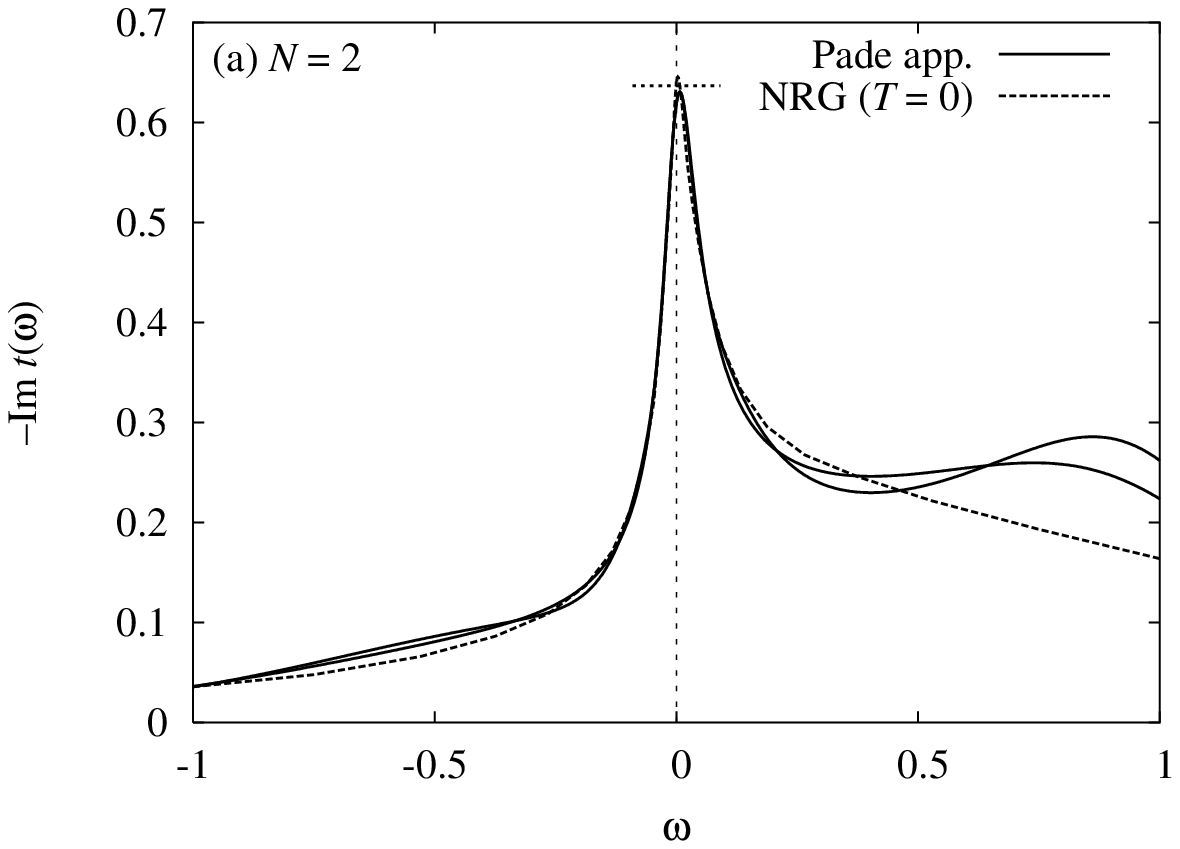}
	\includegraphics[width=7cm]{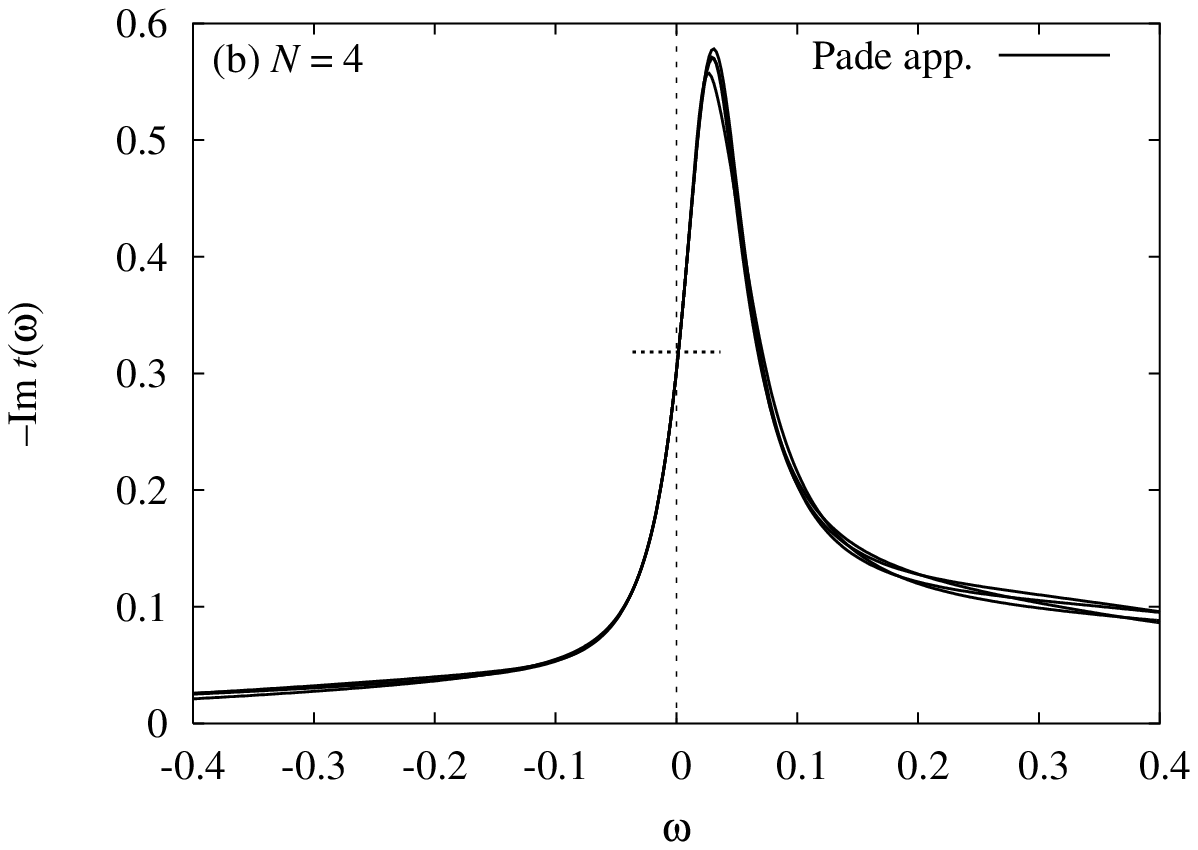}
	\includegraphics[width=7cm]{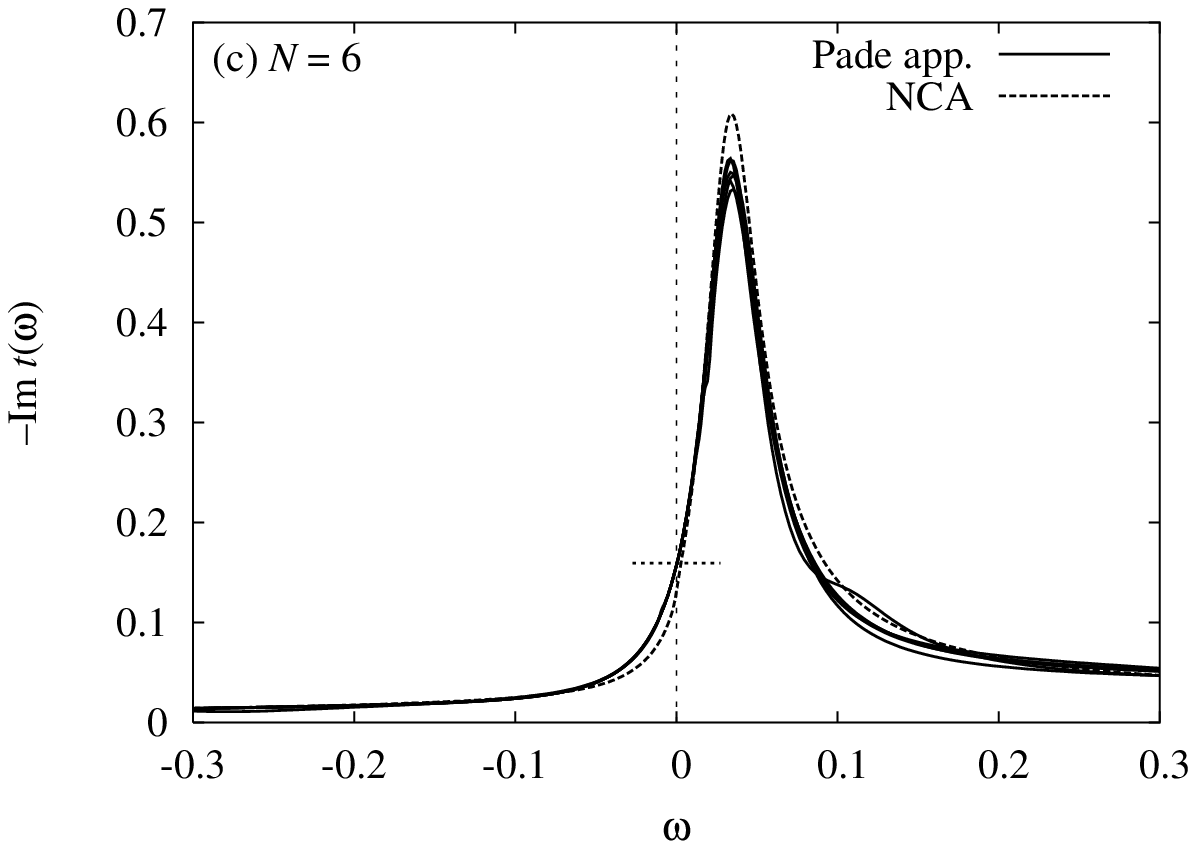}
	\includegraphics[width=7cm]{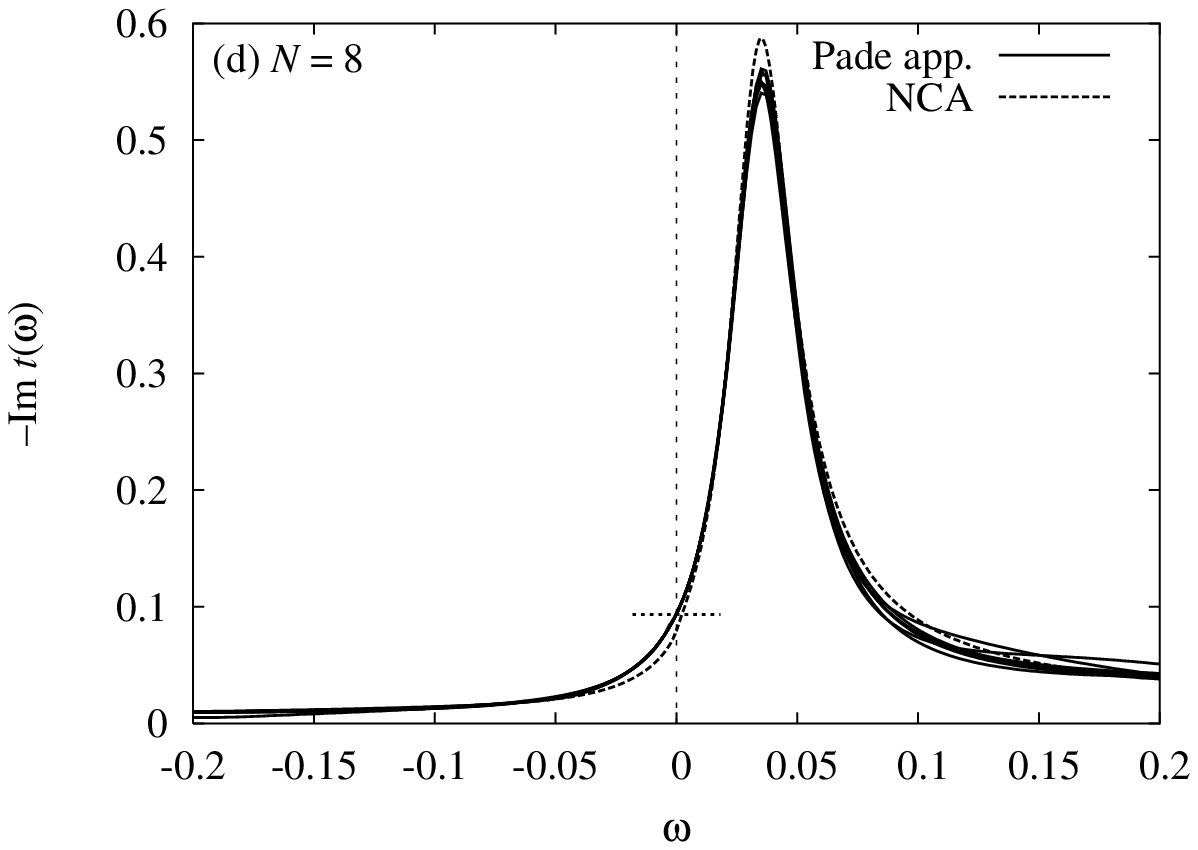}
	\end{center}
	\caption{The $t$-matrix $-\text{Im} t(\omega+{\rm i}\delta)$ computed in the Pad\'e approximation at $T=0.001$ and comparison with the NRG ($N=2$) and the NCA ($N=6$ and 8). Parameters are listed in Table~\ref{tab:param}. Marks at $\omega=0$ indicate exact values obtained from the Friedel sum rule, eq.~(\ref{eq:friedel}). }
	\label{fig:n2468-t_pade}
\end{figure}
Figure~\ref{fig:n2468-t_pade} shows $-\text{Im}t(\omega)$ computed from the Pad\'e approximation at $T=0.001$ for the parameters listed in Table~\ref{tab:param}. 
We plot $N$ components separately to check the accuracy of the analytic continuations. 
We find an excellent agreement between all components around the Fermi level,  while at higher energies the spectra show some differences. 
The extent of the difference can be regarded as a measure of the error in the approximation. 
For comparison, we plot results computed using the numerical renormalization group (NRG)\cite{Wilson, Sakai} at $T=0$ for $N=2$, and results from the NCA for $N=6$ and 8. 
It turns out that for all $N$ the overall shapes obtained by the different methods are in good agreement. 
\begin{figure}[tb]
	\begin{center}
	\includegraphics[width=8cm]{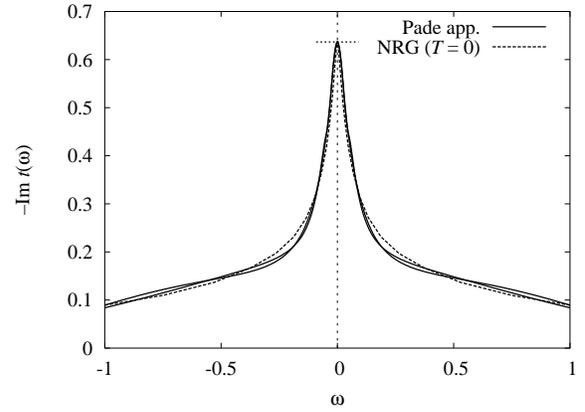}
	\end{center}
	\caption{The $t$-matrix $-\text{Im} t(\omega+{\rm i}\delta)$ of the Kondo model computed in the Pad\'e approximation. Parameters are $J=0.3$ and $T=0.001$. The dashed line is the NRG result at $T=0$. }
	\label{fig:kondo-t_pade}
\end{figure}
Results for the Kondo model are shown in Fig.~\ref{fig:kondo-t_pade}. 
In computing the spectrum, we have set the real part of $t({\rm i}\epsilon_n)$ to 0 neglecting tiny statistical errors. 
We have confirmed that the spectral function is symmetic with respect to $\omega=0$ and agrees with the result from NRG.

At $T=0$ the Friedel sum rule relates the $t$-matrix at the Fermi level with the occupation number of the local state. 
For the constant density of states represented in eq.~(\ref{eq:dos_c}), $-\text{Im} t_{\alpha}(0+{\rm i}\delta)$ is given by\cite{Hewson}
\begin{align}
	-\text{Im} t_{\alpha}(0+{\rm i}\delta) = \frac{1}{\pi \rho_0} \sin^2 (\pi n_{\alpha}),
\label{eq:friedel}
\end{align}
where $n_{\alpha}= \langle X_{\alpha \alpha} \rangle$ is the mean occupation of the state $\alpha$.
The exact values of $-\text{Im} t(0+{\rm i}\delta)$ have been marked in Figs.~\ref{fig:n2468-t_pade} and \ref{fig:kondo-t_pade}. 
We remark that our results computed in the Pad\'e approximation agree with the Friedel sum rule for all $N$. 
We conclude that analytic continuation by the Pad\'e approximation works for the CT-QMC data. 
In the case where there is an energy gap or several excitations in the spectrum, however, the Pad\'e approximation may not provide this level of accuracy.

\subsection{Dynamical susceptibility}

We next present spectra of two-particle response functions. 
\begin{figure}[tb]
	\begin{center}
	\includegraphics[width=7cm]{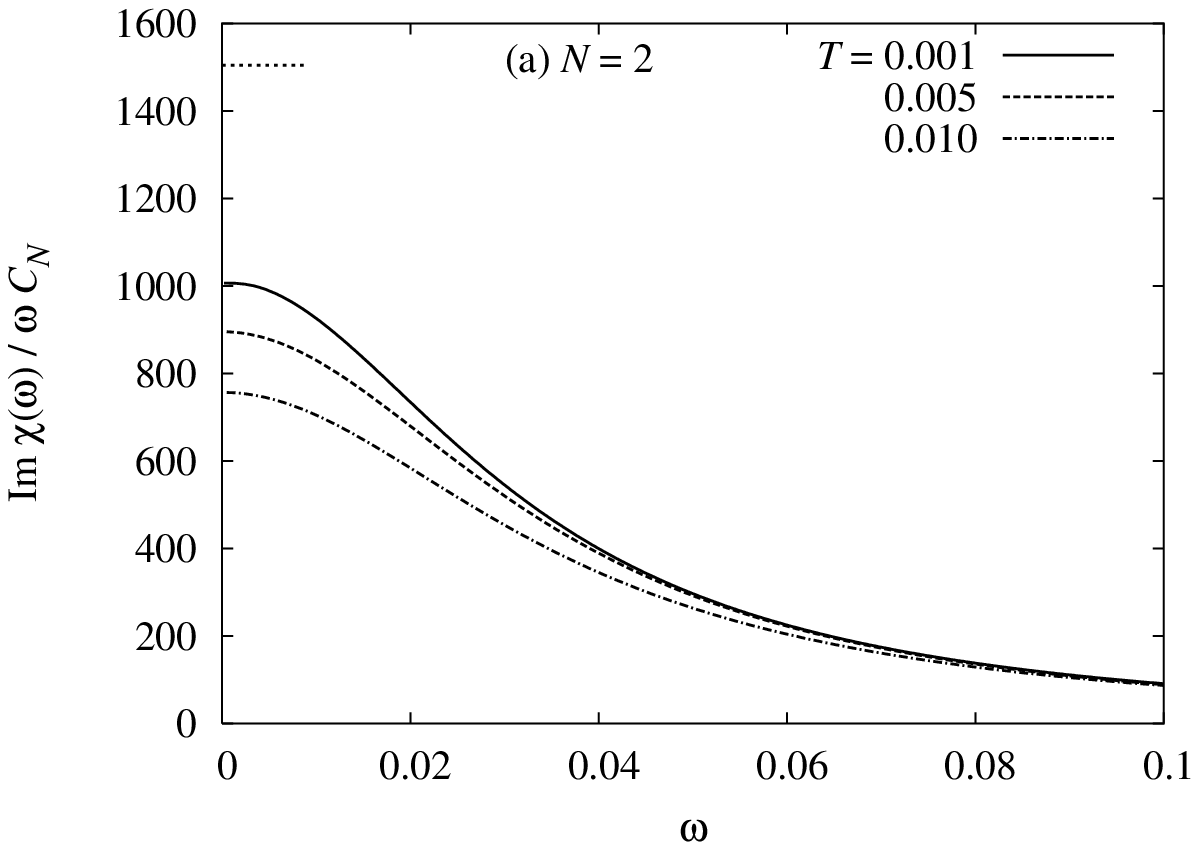}
	\includegraphics[width=7cm]{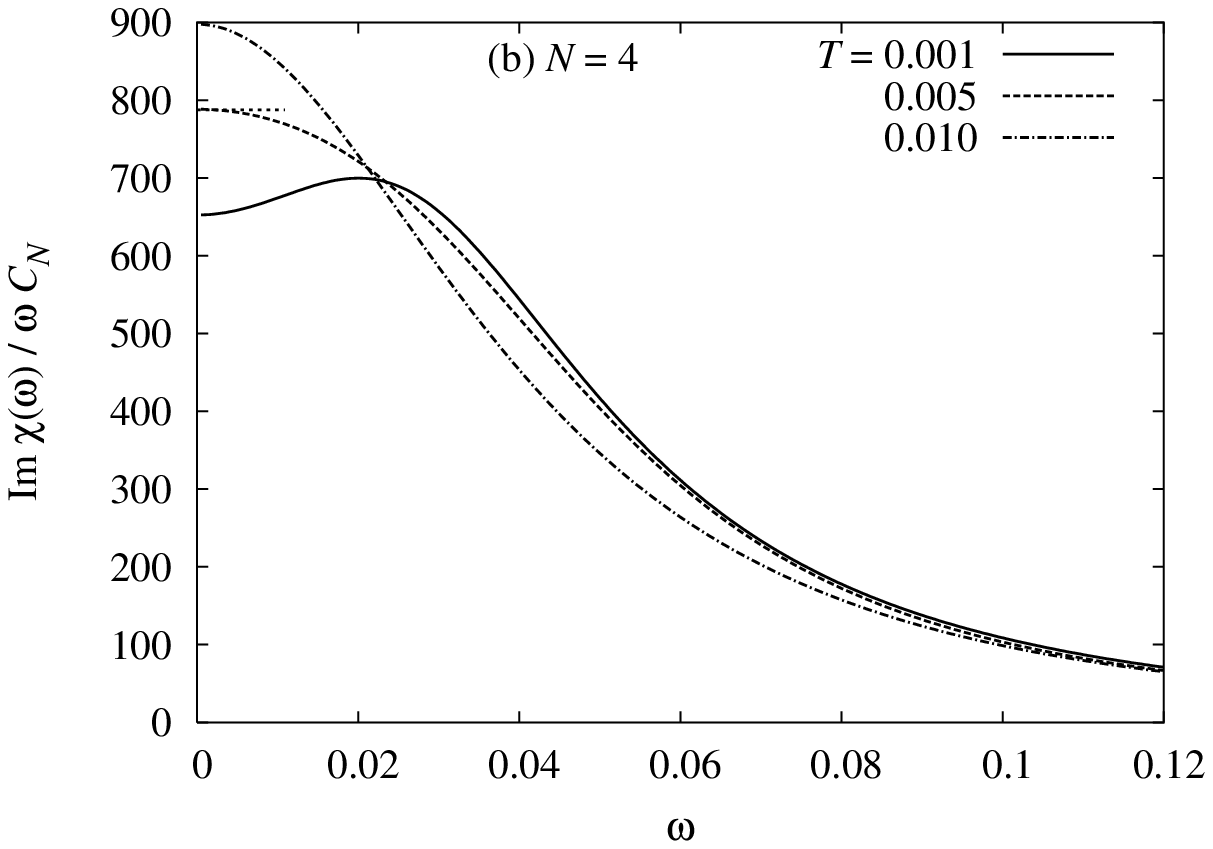}
	\includegraphics[width=7cm]{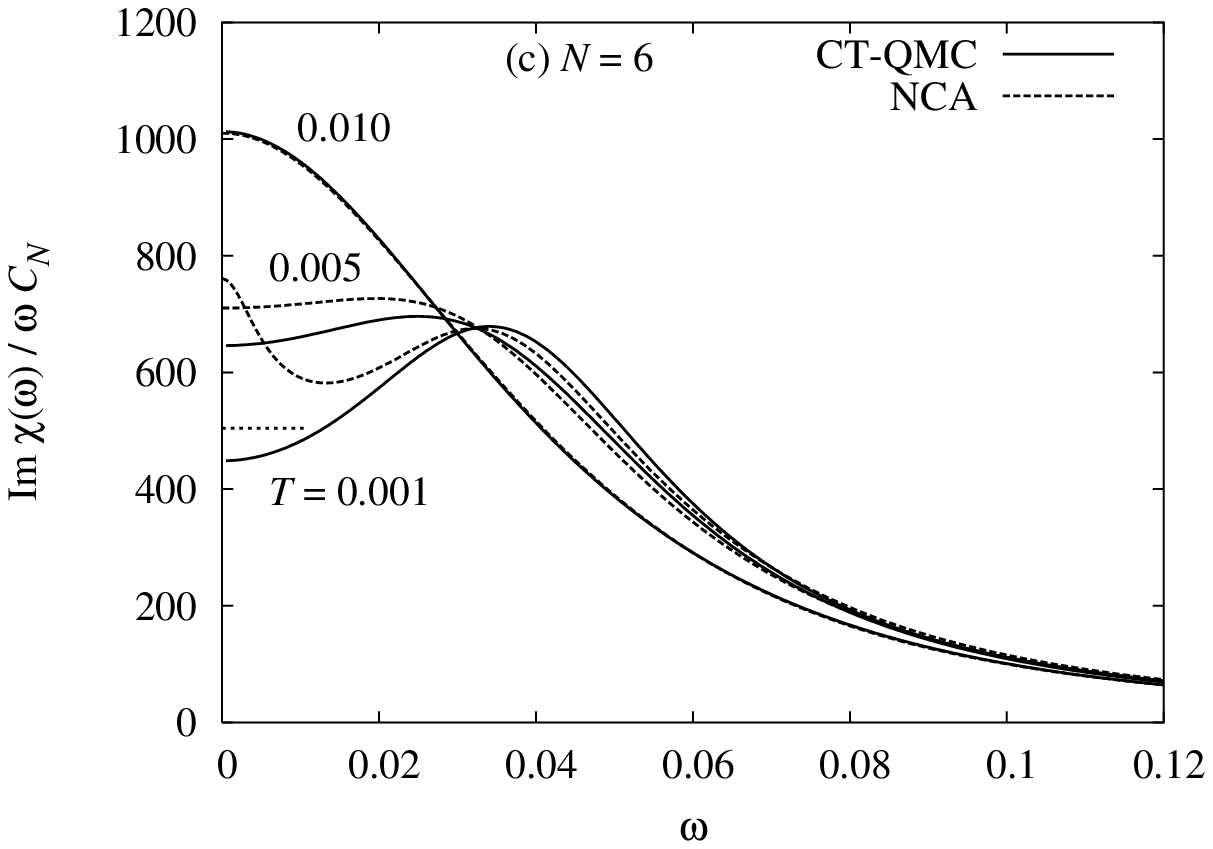}
	\includegraphics[width=7cm]{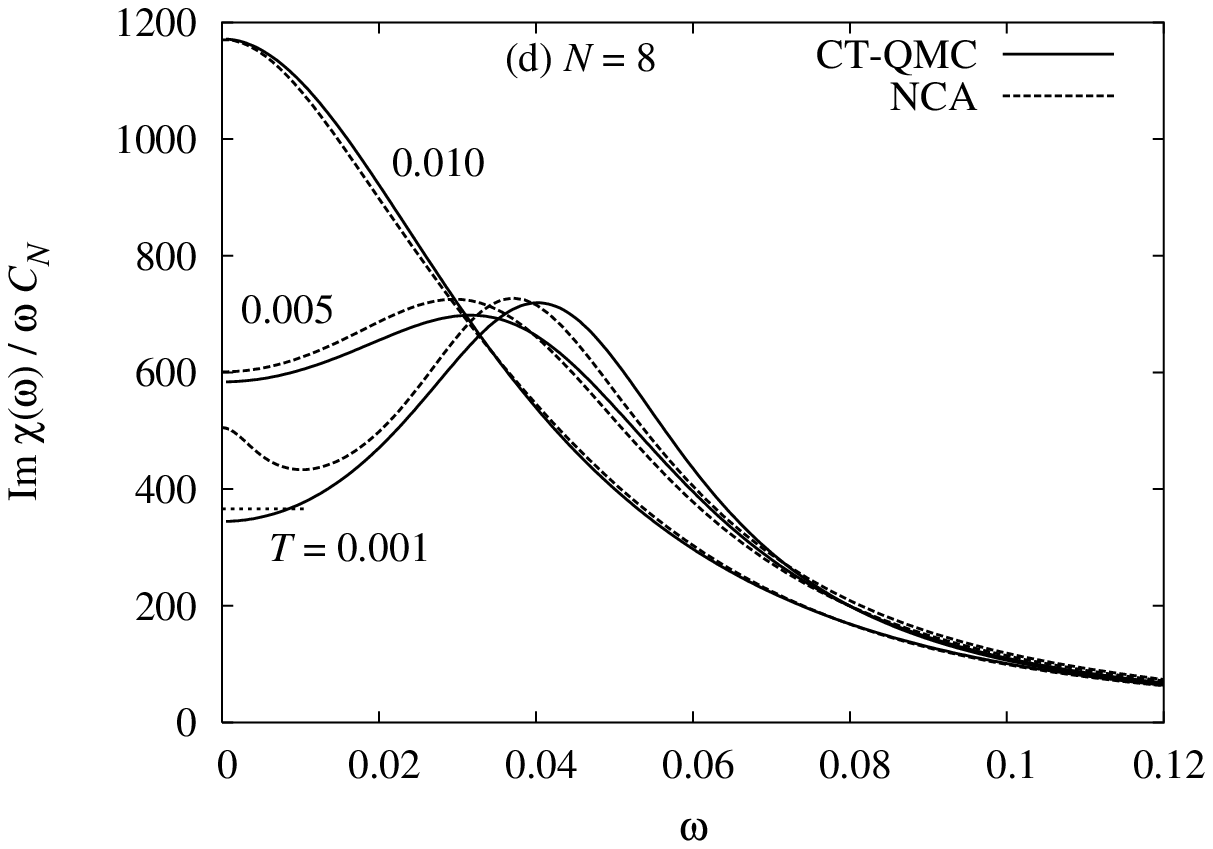}
	\end{center}
	\caption{$\text{Im}\chi(\omega)/\omega$ for several temperatures. Parameters are listed in Table~\ref{tab:param}. Marks at $\omega=0$ indicate values deduced from the Korringa-Shiba relation, eq.~(\ref{eq:ksr}), with $a=1$ and $\chi(T=0.001)$. For $N=6$ and 8, results from the NCA are plotted by a dashed line.}
	\label{fig:n2468-chi_pade}
\end{figure}
Figure~\ref{fig:n2468-chi_pade} shows $\text{Im}\chi(\omega +{\rm i}\delta)/\omega$ computed in the Pad\'e approximation for the same parameters as in Fig.~\ref{fig:n2468-t_pade}. 
Since a comparison between $N$ components, which is made in Fig.~\ref{fig:n2468-t_pade}, is not available in this case, we have confirmed reproducibility of the spectra between different Monte Carlo ensembles. 
Spectra for $N=2$ are almost Lorentzian for all plotted temperatures, while for $N \geq 4$ a peak appears at a finite energy at low temperatures. 
This is consistent with the $t$-matrix in Fig.~\ref{fig:n2468-t_pade}, where the Kondo resonance shifts to higher energy with increasing $N$. 
We also plot, for reference, results computed in the NCA for $N=6$ and 8.
In Figs.~\ref{fig:n2468-chi_pade}(c) and (d), we can see excellent agreement between two results at $T=0.01$, which is of the order of $T_{\rm K}$.  
At lower temperatures, on the other hand, both results do not agree around $\omega=0$ due to the unphysical peak of the NCA\cite{nca3}.

In the Fermi liquid regime, the Korringa-Shiba relation (KSR) connects the imaginary part of the dynamical susceptibility with the static ones. \cite{Hewson,Shiba}
It has been proven for the orbitally degenerate Anderson model in the wide band limit. 
In models with finite band width, however, the static susceptibility is enhanced and does not satisfy the KSR. 
This is due to the fact that the cutoff of the conduction band enhances the static susceptibility over the universal value, while it does not affect the low-energy imaginary part. 
Appendix B demonstrates the deviation for the non-interacting Anderson model. 
To account for deviations from the wide band limit, we introduce a parameter $a$ and modify the original KSR as
\begin{align}
	\lim_{\omega \rightarrow 0} \frac{a^2}{\chi^2} \frac{\text{Im} \chi(\omega +{\rm i}\delta)}{\omega}
	 = \frac{\pi}{N C_N}.
\label{eq:ksr}
\end{align}
Here $a=1$ corresponds to the KSR, but $a$ is greater than unity in the case of finite band width.
We have marked, in Fig.~\ref{fig:n2468-chi_pade}, values of $\text{Im}\chi(\omega+{\rm i}\delta)/\omega|_{\omega=0}$ deduced from $\chi$ at $T=0.001$ and $a=1$ in eq.~(\ref{eq:ksr}). 
Comparing with spectra at $T=0.001$, we observe large deviations for small $N$. 
For $N=2$, $\chi$ with $a=1$ leads to about 50\% deviation from the actual value of $\text{Im}\chi(\omega+{\rm i}\delta)/\omega|_{\omega=0}$.
Since the width of the Kondo resonance is not negligible as compared with $D$ especially for $N=2$, as shown in Fig.~\ref{fig:n2468-t_pade}, the effect of the finite cutoff cannot be neglected. 

\begin{figure}[tb]
	\begin{center}
	\includegraphics[width=8cm]{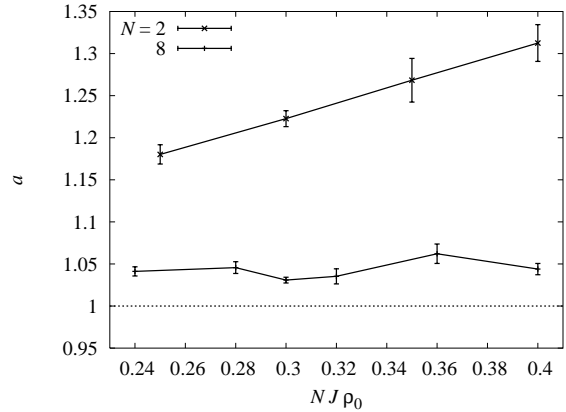}
	\end{center}
	\caption{The parameter $a$ in eq.~(\ref{eq:ksr}) as a function of $NJ\rho_0$ for $N=2$ and 8.}
	\label{fig:ksr}
\end{figure}
To see the validity of the KSR in the wide band limit, or equivalently, in the weak coupling limit, we plot the parameter $a$ against $NJ\rho_0$ in Fig.~\ref{fig:ksr}. 
In this figure, error bars have been estimated from the statistical errors of $\chi$. 
We note that the errors actually come from inaccuracies of $\text{Im}\chi(\omega+ {\rm i}\delta)$ as well due to the analytic continuation. 
For $N=8$, the parameter $a$ deviates from unity only by about 5\%. 
For $N=2$, on the other hand, $a$ varies linearly against $NJ\rho_0$. 
The deviation from unity is almost proportional to $NJ\rho_0$ in the range of our Monte Carlo simulations.
Consequently, in order to satisty the KSR, $NJ\rho_0$ should be much smaller than unity for $N=2$.

\section{Summary}

We have extended the CT-QMC method for fermions to the Coqblin-Schrieffer model.
An arbitrary number $N$ of local states can be treated by our algorithm 
and for antiferromagnetic interactions, the scheme is free from the minus sign problem. 
Therefore it is a powerful tool to study heavy fermion systems within the framework of DMFT.

We have computed spectral functions using the Pad\'e approximation for the analytic continuation.
The impurity $t$-matrix shows an excellent correspondence between the $N$ components at low frequencies, which demonstrates the good quality of the imaginary-time data obtained using the new scheme. 
The accuracy of the magnetic spectra has been examined by comparison with the NCA for large $N$. 
Both results are in good agreement at temperatures of the order of $T_{\rm K}$. 
At lower temperatures, where the NCA cannot accurately reproduce the low-energy excitations, the present scheme still works properly.

The spectra have been tested using available Fermi liquid relations. 
We have found deviations from the Korringa-Shiba relation, whereas the $t$-matrix satisfies the Friedel sum rule for all $N$. 
The deviations are due to the finite cutoff of the conduction band, which affects the static susceptibility but not the low-energy spectrum. 
These results reveal a weak point of the CT-QMC: it is difficult to approach the universal regime with regard to the static susceptibility.

The present algorithm can easily be applied to DMFT simulations of the Coqblin-Schrieffer lattice model. 
Namely, the effective medium is optimized with use of the impurity $t$-matrix. 
It is possible to address the formation of heavy quasi-particles through the conduction electron Green function as well as the $t$-matrix of the lattice systems.

\section*{Acknowledgment}
One of the authors (J. O.) is supported by Research Fellowships of the Japan Society for the Promotion of Science for Young Scientists.

\appendix

\section{Derivation of update probabilities in the CS model from the Anderson model}
The CS model is derived from the Anderson model by taking the localized limit with strong correlations\cite{CS}. 
Namely, charge fluctuations are suppressed by infinite Coulomb repulsion and the deep local level. 
We take the limits $\epsilon_f \rightarrow -\infty$ and $V^2 \rightarrow \infty$ with $J=-V^2/\epsilon_f$ fixed, where $\epsilon_f$ and $V$ denote the energy of the local level and strength of hybridization, respectively. 
Similarly, update probabilities in the CS model can be derived from the corresponding expressions in the Anderson model. 
In this appendix, we demonstrate the way to take the limit in the Monte Carlo formalism.

We consider the Anderson model of spin-less fermions for simplicity. 
In this case, the localized limit leads to a potential scattering, or equivalently, the CS model with $N=1$.
The update probability for cutting a segment (addition of an anti-segment) is given by\cite{Werner}
\begin{align}
	\frac{p(k \rightarrow k+1)}{p(k+1 \rightarrow k)}
	= V^2 {\rm e}^{l \epsilon_f} \left( -\frac{\det D^{(+)}}{\det D} \right) \frac{\beta l_{\rm max}}{k+1},
\label{eq:prob_Anderson}
\end{align}
where $l$ denotes a length of a segment which will be removed. 
$D^{(+)}$ is the matrix obtained by adding the operators $c^{\dag}(\tau) c(\tau+l)$ to the end of the matrix $D$. 
In the limit of $\epsilon_f \rightarrow -\infty$, the probability of such an update with $l$ finite becomes 0 due to the factor ${\rm e}^{l\epsilon_f}$. 
Hence $l$ should be reduced as the inverse of $|\epsilon_f|$ in the update.
For this purpose, we introduce a cutoff length $l_0$ defined by
\begin{align}
	l_0 = \frac{\lambda}{-\epsilon_f}, \qquad (\lambda \gg 1).
\end{align}
If $l \geq l_0$, the update is negligible due to the factor ${\rm e}^{-\lambda}$.
Hence we can restrict the length to $l<l_0$.
The restriction for $l$ replaces the factor $l_{\rm max}$ with $l_0$ in eq.~(\ref{eq:prob_Anderson}). 
Taking $l=x l_0$ with $0<x < 1$, the update probability is given in terms of $J$ by
\begin{align}
	\frac{p(k \rightarrow k+1)}{p(k+1 \rightarrow k)}
	= J \lambda {\rm e}^{-\lambda x} \left( -\frac{\det D^{(+)}}{\det D} \right) \frac{\beta}{k+1}.
\end{align}
Since the probability is independent of $x$ in the limit of $\epsilon_f \rightarrow -\infty$, we integrate out $x$ as follows:
\begin{align}
	\int_0^{1} {\rm d}x \lambda {\rm e}^{-\lambda x} \simeq 1.
\end{align}
This equality becomes exact in the limit $\lambda \rightarrow \infty$, which is realized in the limit $\epsilon_f \rightarrow -\infty$. 
As a result, we obtain the update probability that 
the localized electron is removed for an infinitesimal time as follows:
\begin{align}
	\frac{p(k \rightarrow k+1)}{p(k+1 \rightarrow k)}
	= J \left( -\frac{\det D^{(+)}}{\det D} \right) \frac{\beta}{k+1}.
\end{align}
This formula is identical with the probability that the operator $J_{\alpha \alpha} X_{\alpha \alpha} (-c_{\alpha} c_{\alpha}^{\dag})$ is added. 
It is obvious from this derivation that $D^{(+)}$ is the matrix with $c^{\dag}(\tau) c(\tau+0)$ added to the original one, and that the equal-time Green function should be $g(+0)$. 
In a similar manner, all formulae of transition probabilities in the CS model can be derived from the corresponding processes in the Anderson model.

\section{The Korringa-Shiba relation in a model with finite band width}
The Korringa-Shiba relation connects ${\rm Im}\chi(\omega+{\rm i}\delta) / \omega|_{\omega=0}$ to $\chi^2(0)$ by a universal value\cite{Shiba}. 
The equation has been proven in the wide band limit. 
Hence it may not be satisfied in numerical calculations for systems with finite band width, as in the present study.
In order to clarify the deviation from the universal value, we consider the non-interacting Anderson model with $N$-fold degeneracy. 
Assuming a constant density of states, eq.~(\ref{eq:dos_c}), the Matsubara Green function is given by
\begin{align}
	G_f^{-1} ({\rm i}\epsilon_n)
	&= {\rm i}\epsilon_n - \epsilon_f +(2{\rm i}\Delta /\pi) \arctan ( D/\epsilon_n ) \nonumber \\
	&\simeq {\rm i}\epsilon_n a^{-1} - \epsilon_f + {\rm i}\Delta \text{sgn}(\epsilon_n),
\end{align}
where $\Delta=\pi V^2 \rho_0$ and
\begin{align}
	a^{-1} = 1- \frac{2\Delta}{\pi D}.
\end{align}
We have assumed $\Delta/D \ll 1$. 
The wide band limit corresponds to $a=1$, and a finite band 
produces a correction proportional to $\Delta /D$.

We evaluate the dynamical susceptibility using the above Green function including the effect of the finite band width.
The dynamical susceptibility is given by
\begin{align}
	\chi({\rm i}\nu_n) = -NC_N T \sum_{n'} G_f({\rm i}\epsilon_{n'}) G_f({\rm i}\epsilon_{n'} + {\rm i}\nu_n),
\end{align}
where $\nu_n=2n \pi T$ is the boson Matsubara frequency.
Evaluating the imaginary part at $T=0$, we obtain the well-known relation
${\rm Im}\chi(\omega+{\rm i}\delta) / \omega|_{\omega=0} = \pi NC_N \rho_f^2(0)$.
It turns out that the quantity $a$ does not affect the low-energy spectrum.
On the other hand, the real part is influenced by the cutoff of the band. 
The sum over the Matsubara frequency is replaced by an integral along the imaginary-frequency axis, and evaluates to 
\begin{align}
	\chi(0) = aNC_N \rho_f(0).
\end{align}
Therefore for $a=1$, the Korringa-Shiba relation is satisfied.
On the other hand, if $a>1$, the static susceptibility is enhanced, and the Korringa-Shiba relation does not hold. 
In the Anderson model with $U \neq 0$ or in the Coqblin-Schrieffer model, $\Delta$ is replaced by the width of the Kondo resonance, and therefore is of the order of the Kondo temperature.


\begin{thebibliography}{99}
\bibitem{Georges} A. Georges, G. Kotliar, W. Krauth and M. J. Rozenberg: Rev. Mod. Phys. \textbf{68} (1996) 13. 
\bibitem{Rubtsov} A.N. Rubtsov, V.V. Savkin and A.I. Lichtenstein: Phys. Rev. B \textbf{72} (2005) 035122.
\bibitem{Werner} P. Werner, A. Comanac, L.de' Medici, M. Troyer and A.J. Millis: Phys. Rev. Lett. \textbf{97} (2006) 076405; 
P. Werner and A.J. Millis: Phys. Rev. B \textbf{74} (2006) 155107.
\bibitem{Yoo} The absence of a sign problem can be demonstrated in analogy to the proof for the Hirsch-Fye method in Yoo et al., J. Phys. A: Math. Gen. \textbf{38} (2005) 10307. We thank R. Kaul for bringing this to our attention.
\bibitem{Gull} E. Gull, P. Werner, A.J. Millis and M. Troyer: cond-mat/0609438. 
\bibitem{Werner_doping} P. Werner and A.J. Millis: Phys. Rev. B \textbf{75} (2007) 085108. 
\bibitem{Haule} K. Haule: Phys. Rev. B \textbf{75} (2007) 155113. 
\bibitem{Werner_phonon} P. Werner and A.J. Millis: cond-mat/0701730. 
\bibitem{CS} B. Coqblin and J.R. Schrieffer: Phys. Rev. \textbf{185} (1969) 847. 
\bibitem{Fetter-Walecka} A.L. Fetter and J.D. Walecka: \textit{Quantum Theory of Many-Particle Systems} (McGraw-Hill, New York, 1971).

\bibitem{nca1} Y. Kuramoto: Z. Phys. B \textbf{53} (1983) 37; H. Kojima, Y. Kuramoto and M. Tachiki: Z. Phys. B \textbf{54} (1984) 293.
\bibitem{Bickers} N.E. Bickers: Rev. Mod. Phys. \textbf{59} (1987) 845.
\bibitem{nca3} Y. Kuramoto and H. Kojima: Z. Phys. B \textbf{57} (1984) 95.
\bibitem{Otsuki_thermo} J. Otsuki, H. Kusunose and Y. Kuramoto: J. Phys. Soc. Jpn. \textbf{75} (2006) Suppl. 256.

\bibitem{Rajan} V.T. Rajan: Phys. Rev. Lett. \textbf{51} (1983) 308.

\bibitem{pade} H.J. Vidberg and J.W. Serene: J. Low Temp. Phys. \textbf{29} (1977) 179. 
\bibitem{Jarrell-Gubernatis} M. Jarrell and J.E. Gubernatis: Phys. Rep. \textbf{269} (1996) 133.

\bibitem{Wilson} K.G. Wilson: Rev. Mod. Phys. \textbf{47} (1975) 773.
\bibitem{Sakai} O. Sakai, Y. Shimizu and T. Kasuya: J. Phys. Soc. Jpn. \textbf{58} (1989) 3666. 
\bibitem{Hewson} A.C. Hewson: \textit{The Kondo Problem to Heavy Fermions} (Cambridge University Press, 1993).

\bibitem{Shiba} H. Shiba: Prog. Theor. Phys. \textbf{54} (1975) 967. 

\end{thebibliography}
\end{document}